\documentclass[12pt,preprint]{aastex631}

\usepackage{longtable}
\usepackage{lipsum} 
\usepackage{multirow}
\usepackage{natbib}

\usepackage{epsfig}
\usepackage{graphicx}
\usepackage{bm}
\usepackage{amsmath}
\usepackage{amsfonts}
\usepackage{amssymb}
\usepackage{xcolor}
\usepackage{mathrsfs}
\usepackage[caption=false]{subfig}
\usepackage{CJK}
\usepackage{ulem}

\def\arcsec{$^{\prime\prime}$}
\def\arcmin{$^{\prime}$}

\newcommand{\NHthree}{NH\ensuremath{_{3}}}
\newcommand{\NtwoH}{N\ensuremath{_{2}}H\ensuremath{^{+}}}

\def\deg{\ensuremath{^{\circ}}}
\newcommand{\Msun}{\mbox{\ensuremath{{M}_{\sun}}}}
\newcommand{\Lsun}{\mbox{\ensuremath{{L}_{\sun}}}}

\newcommand{\kms}{km~s\ensuremath{^{-1}}}

\newcommand{\cmm}{cm\ensuremath{^{-2}}}

\begin{document}


\title{ Impact of gravity on changing magnetic field orientations in a sample of massive protostellar clusters observed with ALMA}

\shorttitle{}

\author[0000-0003-2384-6589]{Qizhou Zhang}
\affiliation{Center for Astrophysics $\vert$ Harvard \& Smithsonian, 60 Garden Street, Cambridge, MA, 02138, USA}
\email{qzhang@cfa.harvard.edu}

\author[0000-0002-4774-2998]{Junhao Liu}
\affiliation{National Astronomical Observatory of Japan, 2-21-1 Osawa, Mitaka, Tokyo 181-8588, Japan}

\author[0000-0001-6924-9072]{Lingzhen Zeng}
\affiliation{Center for Astrophysics $\vert$ Harvard \& Smithsonian, 60 Garden Street, Cambridge, MA, 02138, USA}
\author[0000-0002-0294-4465]{J. D. Soler}
\affiliation{Istituto di Astrofisica e Planetologia Spaziali (IAPS). INAF. Via Fosso del Cavaliere 100, 00133 Roma, Italy}

\author[0000-0002-9774-1846]{Huei-Ru Vivien Chen}
\affiliation{Institute of Astronomy and Department of Physics, National Tsing Hua University, Hsinchu 30013, Taiwan, ROC}

\author[0000-0001-8516-2532]{Tao-Chung Ching}
\affiliation{National Radio Astronomy Observatory, P.O. Box O, Socorro, NM 87801, USA}

\author[0000-0002-3412-4306]{Paul T. P. Ho}
\affiliation{East Asian Observatory, 660 N. A‘ohōkū Place, University Park, Hilo, HI 96720, USA}
\affiliation{Institute of Astronomy and Astrophysics, Academia Sinica, 11F of Astronomy-Mathematics Building, AS/NTU No.1, Sec. 4, Roosevelt Road, Taipei 10617, Taiwan, ROC}

\author[0000-0002-3829-5591]{Josep Miquel Girart}
\affiliation{Institut de Ciències de l’Espai (ICE, CSIC), Can Magrans s/n, E-08193 Cerdanyola del Vallès, Catalonia, Spain}
\affiliation{Institut d’Estudis Espacials de de Catalunya (IEEC), E-08034 Barcelona, Catalonia, Spain}

\author[0000-0003-2777-5861]{Patrick M. Koch}
\affiliation{Institute of Astronomy and Astrophysics, Academia Sinica, 11F of Astronomy-Mathematics Building, AS/NTU No.1, Sec. 4, Roosevelt Road, Taipei 10617, Taiwan, ROC}

\author[0000-0001-5522-486X]{Shih-Ping Lai}
\affiliation{Institute of Astronomy and Department of Physics, National Tsing Hua University, Hsinchu 30013, Taiwan, ROC}

\author[0000-0003-1275-5251]{Shanghuo Li}
\affiliation{School of Astronomy and Space Science, Nanjing University, 163 Xianlin Avenue, Nanjing 210023, Jiangsu, People's Republic of China}

\author[0000-0002-7402-6487]{Zhi-Yun Li}
\affiliation{Astronomy Department, University of Virginia, Charlottesville, VA 22904-4325, USA}

\author[0000-0003-2300-2626]{Hauyu Baobab Liu}
\affiliation{Department of Physics, National Sun Yat-Sen University, No. 70, Lien-Hai Road, Kaohsiung City 80424, Taiwan, ROC}
\affiliation{Center of Astronomy and Gravitation, National Taiwan Normal University, Taipei 116, Taiwan, ROC}

\author[0000-0002-5093-5088]{Keping Qiu}
\affiliation{School of Astronomy and Space Science, Nanjing University, 163 Xianlin Avenue, Nanjing 210023, Jiangsu, People's Republic of China}
\affiliation{Key Laboratory of Modern Astronomy and Astrophysics (Nanjing University), Ministry of Education, Nanjing 210023, Jiangsu, People's Republic of China}

\author[0000-0002-1407-7944]{Ramprasad Rao}
\affiliation{Center for Astrophysics $\vert$ Harvard \& Smithsonian, 60 Garden Street, Cambridge, MA, 02138, USA}

\begin{abstract}
The magnetic field is integral to our understanding of the formation and dynamical evolution of molecular clouds and star formation within. 
We present a polarimetric survey of 17 massive protostellar cluster forming clumps, covered in 34 pointings in the 230-GHz window using the Atacama Large Millimeter/submillimeter Array (ALMA). The two array configurations, C43-1 and C43-4, probe linearly polarized dust emission, hence the plane-of-the-sky orientation of magnetic fields, at resolutions of 1\arcsec\ and 0\arcsec.4 that correspond to approximately 0.01pc core and $10^3$ au envelope scales, respectively. The relative orientations (ROs) of the magnetic field probed at two spatial scales  are analyzed for the entire protostellar cluster sample and for a subset of objects in NGC 6334. We found a bimodal distribution of ROs with peaks at 0\deg\ (parallel) and 90\deg (orthogonal) for the entire sample combined as well as for NGC 6334. We investigate the physical origin of this bimodal distribution through a projected Rayleigh statistic (PRS) analysis in relation to column densities and local gravity in NGC 6334. We found an excess of parallel magnetic fields at column densities $> 10^{23}$ \cmm. The underlying cause of the RO distribution of the magnetic field is gravitational collapse at higher gas densities, which drags and reorients the magnetic field as shown in the alignment between the magnetic field and the direction of gravitational forces. The distribution of ROs observed here is consistent with the evolution of relative orientations of an initially sub-Alv\'enic cloud that becomes magnetically super-critical and super-Alv\'enic as the cloud collapses to form stars. 
\end{abstract}


\section{Introduction}
\label{sec:intro}

The formation of stars involves a significant increase in densities in molecular clouds as materials condense and collapse under gravity and continue to accrete toward protostars. Strongly coupled with the molecular gas in interstellar medium, the magnetic field restricts gas motions across the field lines, thus, hinders the mass accretion. Therefore, the presence of strong magnetic fields can prolong the time scale of star formation and reduce its efficiency \citep[e.g.,][]{hennebelle2019}. 

Although the magnetic field is one of the central components in many theories of star formation \citep[e.g.,][]{shu1987}, its direct measurements have been limited due to a lack of sensitivity even with the most powerful telescopes. Zeeman splitting of molecular lines due to interactions of magnetic fields and the magnetic moment of the molecule is the most direct technique to measure the line-of-sight component of the field in the interstellar medium. However, the splitting due to the typical field strength in a cloud is smaller than the spectral linewidth. Therefore, it severally limits the application in molecular clouds. The reader is referred to \citet{crutcher2012} for a review on this topic. A more viable alternative to Zeeman observations is to measure the linearly polarized emission from dust grains, or spectral lines due to the Goldreich-Kylafis effect \citep{goldreich1981}, which probes the topology of the plane-of-the-sky component of the magnetic field. Although polarimetric observations do not directly measure the field strength as the Zeeman effect does, this technique can be revealing through a variety of analyses, e.g., relative orientations (ROs) of magnetic fields at different spatial scales \citep{zhang2014, hull2019}, relative orientations of magnetic fields with respect to gas densities \citep{koch2012a, soler2013}, and relative orientations of magnetic fields with respect to turbulence \citep{lazarian2018}. In addition, quantitative analyses can provide estimates of the plane-of-the-sky component of the magnetic field strength \citep{davis1951a, chandra1953a, hildebrand2009, houde2016}. The reader is referred to \citet{liu_J2022b} for a review of the analysis methods and their limitations. The study in this paper focuses on RO analyses of magnetic fields between the core and envelope scales.

Telescopes equipped with polarimeters continue to push the frontiers in mapping magnetic fields in interstellar medium and molecular clouds at multiple spatial scales. Throughout the paper, we use {\it clump} to refer to parsec-scale molecular structures with masses of approximately 10$^3$ \Msun\ that are capable to form a small cluster of stars; {\it core} to refer to 0.01 to 0.1 pc gas structures that are capable of forming one to multiple stars; and {\it protostellar envelope} to refer to molecular structures of 10$^3$ au surrounding protostars. Observations of polarized emission in star forming clouds were attempted routinely by JCMT and CSO, and more recently by SOFIA and IRAM 30m Telescope at resolutions on the order of $10''$. The Berkeley, Illinois and Maryland Array (BIMA) pioneered polarimetric observations at arcsecond resolutions in the millimeter wavelength. Readers are referred to recent reviews by \citet{pattle2019a, hull2019, pattle2022a} for studies of magnetic fields with both single-dish telescopes and radio interferometers. On the subject of high-mass star formation, parsec-scale clumps harboring high-mass protostars were among the first star forming regions imaged at arcsecond resolutions \citep[e.g.,][]{lai2001}, thanks to their strong dust continuum emission. Subsequently, CARMA, SMA and ALMA have extended full polarization observations to a large number of clumps \citep[e.g.,][]{girart2009, hull2013, girart2013, li_H2015, qiu2013, qiu2014, sridharan2014, cortes2008, cortes2021, sanhueza2021, beuther2024}. These studies revealed a wide range of magnetic field topologies, ranging from hourglass to spiral, and to a more randomly distributed field geometry. Based on the small fraction of sources with hourglass-shaped magnetic fields, \citet{hull2019} concluded that the magnetic field does not dominate the dynamical process during high-mass star formation, but it can be important relative to turbulence \citep[e.g.,][]{girart2013, ching2016}.

While studies of individual high-mass star forming clumps are valuable, it is crucial to map a large sample of clumps to gain a statistical insight into the ensemble behavior of magnetic fields in high-mass star formation. 
\citet{koch2014} analyzed the polarization data of 50 low- and high-mass star forming regions obtained with the SMA and CSO and found that magnetic fields are important relative to turbulence. \citet{liu_J2022b} compiled a sample of low- and high-mass star forming regions with dust polarization observations in the literature. Applying the correction factors derived in \citet{liu_J2022a} to the Davis-Chandrasekhar-Fermi (DCF) method, they found an increasing trend in the mass-to-magnetic-flux ratio with increasing column densities. In addition, the mass-to-flux ratio transitions from $< 1$ to $> 1$ relative to the critical value at a column density of $10^{21}$ \cmm.
With the caveat that the data were collected from a mixed sample of both low- and high-mass star forming regions and that they may be subject to varying environmental influences, such a trend is consistent with the expectation that gravity becomes more dominant in regimes of high mass densities. The transition density at which the mass-to-flux ratio reaches unity is similar to that derived from Zeeman observations \citep{crutcher2012}. These studies indicate that during the evolution of molecular clouds, the dynamical role of magnetic fields relative to gravity and turbulence will evolve. 

Another tool for assessing the dynamic role of magnetic fields is to compare their relative orientations measured at different spatial scales. 
\citet{zhang2014} carried out full polarization observations of 14 massive protostellar cluster forming clumps with the SMA. By measuring relative orientations of magnetic fields in the parsec-scale molecular clumps probed by single-dish telescopes at a resolution of $\sim 10''$  with those probed by the SMA at arcsecond resolutions (corresponding to $<$ 0.1 pc core scales), \citet{zhang2014} found that fields are correlated between these two scales. This finding indicates that the magnetic field is dynamically important during protostellar cluster formation as molecular clumps collapse and fragment. In addition, the study found that the magnetic field in cores does not correlate with the outflow axis. Notwithstanding the small number of outflows in the statistics, these findings suggest the presence of a spatial scale between 0.1~pc cores to $10^2$~au disks, at which magnetic fields are overtaken by other effects such as angular momentum in the gas.

Motivated by the SMA study in \citet{zhang2014}, we conducted ALMA full polarization observations of 17 protostellar cluster forming clumps at even higher angular resolutions to probe magnetic fields at envelope (10$^3$ au) scales. The SMA study suggests that between the protostellar core scale of 0.01 - 0.1 pc, and the accretion disk scale of $10^2$~au, the energy from gravity and angular momentum significantly exceeds the magnetic field energy. The goal of this ALMA study is to image the magnetic field at protostellar envelope scale of $10^3$ au.
The ALMA sample overlaps considerably with the objects in the SMA study by \citet{zhang2014} except those in the Cygnus-X molecular cloud. At a declination of $> 40^\circ$, these Cygnus-X clumps are challenging to observe with ALMA due to their low elevations. In addition, we included two infrared dark clouds G28.34+0.06 and G14.225-0.506 to investigate magnetic fields at an earlier stage of high-mass star formation. This paper reports observations of the entire sample and a multi scale analysis of magnetic fields observed at $\sim 1$\arcsec\ and 0\arcsec.4 resolutions to assess the role of magnetic fields at scales from 0.01pc to $10^3$~au.  Detail studies of individual objects have been reported in separate publications~\citep[e.g.,][]{liu_J2020, liu_J2023a} with additional publications to follow. The paper is organized as follows. Section~\ref{sec:obs} describes observations, data calibration and imaging of the sample. Section~\ref{sec:results} presents magnetic field images and analysis. Section~\ref{sec:discussion} discusses the implications of the results and Section~\ref{sec:sum} summarizes the main findings of the study.

\section{Observations}
\label{sec:obs}
 
Observations of 17 high-mass protocluster forming clumps were conducted during 2018 in full polarization mode using the main array of the Atacama Large Millimeter/submillimeter Array (hereafter ALMA) under project 2017.1.00793.S (PI Qizhou Zhang). In order to sample a wide range of spatial scales from 6\arcsec\ to 0\arcsec.4, we adopted array configurations of C43-1 and C43-4. The date of observations, the array configuration and calibrators used for each scheduling block are listed in Table~\ref{tab:1}. The correlator was configured to cover three 1.875 GHz spectral windows centered at sky frequencies of 216.44, 218.54 and 233.55 GHz, respectively. Each spectral window contains 1920 channels, yielding a channel spacing of 0.98 MHz, approximately 1.3 \kms{} at the observing frequencies. Three additional narrower spectral windows, with a bandwidth of 58.6 MHz and a resolution of 0.24 MHz (0.31 \kms) in each window, were employed to cover the CO 2-1, N$_2$D$^+$ 3-2, and $^{13}$CS 5-4 and OCS 19-18 lines.

The visibility measurement sets were initially calibrated by the ALMA North American ARC using the Common Astronomy Software Applications package (CASA) \citep{mcMullin2007}. The calibrated visibilities of the science targets were further self calibrated using the Stokes $I$ continuum emission from all three 1.875 GHz spectral windows for objects with sufficient signal-to-noise ratios. Self-calibration steps involve visibility phases in two to three interactions. The derived gain solutions were applied to both the continuum and the spectral line data in Stokes $I$ only. Since Stokes $Q$ and $U$ products typically do not have sufficient signal-to-noise ratios, self calibrations using all Stokes products are not feasible \citep{sault1995}.

The fully calibrated visibilities were then imaged in CASA. We averaged the line-free channels in the three 1.875 GHz spectral windows to construct the continuum emission, and ran the `tclean' task using Briggs weighting of the visibilities with a robust parameter of 0.5. The default multiscale parameters were used during deconvolution. Table~\ref{tab:2} lists target names, coordinates of 34 pointings, source LSR velocity, distance, the rms noise, and the synthesized beam size. For some clumps, multiple pointings were used to cover their spatial extent. In these cases, mosaic images were made combining the adjacent pointings.

\section{Results}
\label{sec:results}

\subsection{Spatial distributions of magnetic fields}

With a $1 \sigma$ rms sensitivity of 30 to 70 $\mu$Jy in Stokes {$Q$} and {$U$} images, spatially extended polarized emissions are detected in nearly all objects in the sample. We compute polarization position angles (PA) following the equation
\begin{equation}
\theta = 0.5 \times \arctan\left(\frac{I_U}{I_Q}\right),
\end{equation}
where $\theta$ is the position angle of the electric field measured from north to east, and $I_Q$ and $I_U$ represent Stokes $Q$ and $U$ intensities, respectively. Applying a threshold cutoff of 3$\sigma$, the rms of the noise level, to the Stokes $Q$ and $U$ emission, we obtained the PA of the electric field. The PAs of the magnetic field follow by rotating the electric field by $90^\circ$. For data with high signal to noise ratios, the $1 \sigma$ uncertainty in PA is given by $28.65^\circ \frac{\sigma_p}{I_p}$, where $\sigma_p$ and $I_p$ are noise and signal in the polarized emission $\sqrt{I_Q^2 + I_U^2 - \sigma_p^2}$ \citep{serkowski1974,naghizadeh1993}.

Figure 1 presents the Stokes $I$ emission in contours and the plane-of-the-sky component of magnetic field orientations in line segments. Several clumps were observed with pointings separated by approximately $0.5$\,$\times$\,FWHM of the primary beam of the telescope. For these clumps,  mosaic images were made. The data presented in Figure 1 are from the combined observations of the C43-1 and C43-4 configurations, and images are not corrected by the primary beam of the telescope. This does not affect the PA of magnetic fields since it is derived from the ratio of the Stokes $Q$ and $U$ intensities, which are modulated by the primary beam in an identical manner.

Magnetic fields are detected in nearly all objects and display a range of topologies. NGC 6334 was observed toward four massive clumps along the 10~pc massive filamentary cloud \citep{sandell2000,munoz2007,russeil2013} where active star formation takes place. The overall magnetic fields appear to be in the northwest-southeast orientation perpendicular to the 10~pc filament \citep{arzoumanian2021,liu_J2023a}. The previous SMA observations by \citet{zhang2014} reported an hourglass-shaped magnetic field in NGC 6334 I and In \citep[see also][]{li_H2014}. With higher angular resolutions, the ALMA images (see Fig. 1a) revealed spiral-shaped magnetic fields toward the peak of the Stokes $I$ emission in NGC 6334 I \citep[see also][]{liu_J2023a, cortes2021, cortes2024}. Such magnetic structures, which have been reported in other massive star-forming clumps IRAS~18089-1732 \citep{sanhueza2021}, G~327 \citep{beuther2020}, and Sgr B2 \citep{pan2024}, indicate that the field is dragged by the infalling and accreting material toward the center of the protocluster. For NGC~6334 In, the ALMA observations shown in Fig. 1a extend the SMA image in \citet{zhang2014} to include the southeast region. For NGC~6334 IV, the magnetic field appears to follow a filamentary structure seen in the Stokes $I$ emission (see Fig. 1a).
The magnetic field morphology of NGC~6334 V is similar to that from the SMA \citep{Juarez2017}. 
NGC 6334 IR is an infrared dark cloud clump \citep{li_S2020a} at an earlier evolutionary stage with weaker emission in Stokes $I$. Despite the excellent sensitivity, the polarized emission is detected only toward several Stokes $I$ emission peaks (see Fig. 1c). A detailed analysis of the ALMA full polarization observations of NGC 6334 I/In/IV/V can be found in \citet{liu_J2023a}.

G14.225-0.506 (hereafter G14) is an infrared dark cloud in the M17 star formation complex \citep{povich2016, zhao_Y2025}. The cloud exhibits a network of parallel filaments with an extent of 5~pc revealed in the emission of \NHthree\ and \NtwoH \citep{busquet2011,chen_H2019}. Magnetic fields inferred from polarization observations in the infrared R and H band \citep{santos2016} and in the 0.87 mm continuum emission \citep{anez2020} are found to be perpendicular to the long axes of the filaments, suggesting that the field may play an important dynamical role in filament formation \citep{vanloo2014}. The ALMA data presented here (see Figs. 1b and 1c) offer the first subarcsecond resolution image of magnetic fields in the dense gas.

The full polarization observations of G14 were targeted with a total of 9 pointings, six of which were for G14-South (hereafter G14-S) and G14-North (hereafter G14-N), respectively. The remaining three target isolated peaks in the dust continuum emission. The Stokes $I$ intensities and magnetic field orientations are presented in Figs. 1b and 1c. Spatially extended polarized emission is detected in G14-S and G14-N. G14-N has a relatively uniform magnetic field, while the field in G14-S has a larger PA dispersion. A detailed analysis of the ALMA full polarization observations of G14 will be published in a future paper by {A{\~n}ez-L{\'o}pez} et al. (in preparation).

In addition to NGC~6334 and G14, the polarized emission and magnetic fields are detected in G34.43+0.24 (hereafter G34.43), G35.2-0.74 N (hereafter G35.2N), G28.34$+$0.06 (hereafter G28) and IRAS~18360-0537 (hereafter I18360).

G34.43 is associated with IRAS 18507+0121, originally
identified by \citet{miralles1994} who detected the centimeter continuum and \NHthree\ emission.  \citet{rathborne2005} reported a filamentary structure in the 1.2 mm continuum emission extending over 3 pc, coinciding with the 8.0 $\mu$m dark
cloud revealed in observations with the Spitzer Space Telescope. Subsequent VLA observations revealed a long filament in the \NHthree\ emission \citep{lu2014}. Magnetic fields have been inferred from the full polarization observations with CSO/SHARP \citep{tang2019a} and the SMA \citep{zhang2014}. The ALMA observations cover three pointings along the filament. As shown in Fig. 1d, spatially extended magnetic fields are detected with the field predominantly perpendicular to the main axis of the filament.

G35.2N is associated with a massive star forming region at a distance of 2.19~kpc with a luminosity of $3 \times 10^4$ \Lsun\ \citep{dent1989, sanchez2013d}. Several compact continuum sources have been detected in the (sub)millimeter band with ALMA and the SMA along the filament, indicating that the region is forming a small cluster of stars \citep{qiu2013, zhang_Y_2022}. An outflow is detected in both the CO emission \citep{gibb2003,qiu2013} and the near-IR H$_2$ emission \citep{caratti2015}, which extends more than 1\arcmin\ at a position angle perpendicular to the filament. A precessing radio jet is found to be associated with the (sub)mm continuum peak that likely hosts a proto binary \citep{gibb2003,beltran2016}.  Magnetic fields obtained from the full polarization observations with the SMA at the 1\arcsec.5 resolution are mainly along the filament \citep{qiu2013}. As shown in Fig. 1e, the ALMA full polarization observations reveal substructures in the magnetic field at subarcsecond resolutions. In particular, the magnetic field in some filament sections becomes mostly perpendicular to the main axis of the filament. 

G28 is a well-studied infrared dark cloud (IRDC) complex that
contains a total mass of $10^4$ \Msun\ over a projected length of 4.6 pc \citep{rathborne2006, wang2008}.
Three massive clumps, MM1, MM4 and MM9 with  masses of 1000, 1000, and 500 \Msun, respectively, appear to lie in different evolutionary stages of massive star formation based on their IR appearance, FIR luminosities, gas temperatures, and association with complex organic molecules \citep{rathborne2006, wang2006, zhang2009, zhang2015, kong2019}. We observed 4 poiontings in MM4 and two pointings in MM9, respectively (see Fig. 1f). The data from this project, in combination with a separate ALMA project have been analyzed and reported in \cite{liu_J2020}. More recently, \cite{liu_J2024} presented a multi-scale analysis of magnetic fields, including data from Planck, JCMT and ALMA. They found that the G28 cloud over the 10~pc scale is in a trans- to sub-Alfv\'{e}nic state (Alfv\'{e}nic velocity $<$ turbulent velocity), and dense cores in MM9 are magnetically supercritical (mass-to-magnetic-flux ratios exceeding the critical value) and are in a sub virial state (viral parameters $<$ 1.).

I18360 is a massive star-forming region with a luminosity of $1.2 \times 10^5$ \Lsun\ \citep{molinari1996} at a distance of 6.3~kpc \citep{xu_J_2021}. SMA observations revealed an outflow in the CO and SiO emission associated with a massive rotating protostellar core \citep{qiu2012}. The core appears to be elongated along a direction perpendicular to the rotational axis. Subarcsecond resolution observations resolve the core into two continuum peaks separated by 1\arcsec\ \citep{qiu2012,wu2023}. Magnetic fields from the SMA observations are along the core major axis, perpendicular to the outflow \citep{zhang2014}. The ALMA observations build on the SMA observations, revealing magnetic field morphologies both with higher resolutions and with detections of fainter polarized emission (see Fig. 1f).

\subsection{Relative orientation of magnetic fields between different spatial scales}

Relative orientations (ROs) of magnetic fields at different spatial scales can be used to probe the dynamical role of magnetic fields relative to other forces in molecular clouds \citep[e.g.,][]{zhang2014, planck2016b}. Numerical simulations have shown that in strongly magnetized clouds, magnetic fields remain ordered and aligned with the initial field orientations as the cloud evolves \citep[e.g.,][]{ostriker2001, seifried2011}.
For a weakly magnetized protostellar clump where a group of stars form, an initially uniform magnetic field in the clump can be severely distorted around dense cores as materials are accreted toward the cores. Therefore, the degree of correlation between magnetic fields around dense cores and their parental clump signals the dynamical role of the magnetic field with respect to gravity and turbulence. 
\citet{zhang2014} compared the magnetic field orientation in dense cores at 0.1\,pc, sampled with the SMA, and their parental clumps' field orientations, obtained from single-dish telescopes, and found a bimodal distribution in relative orientations. The nonuniform distribution suggests that the magnetic field plays a dynamical role in the fragmentation of clumps and the mass assembly of dense cores.

The ALMA observations obtained from the two different array configurations enable comparisons of magnetic fields measured at an approximately $1''$ resolution in the C43-1 configuration and a 0.4\arcsec\ resolution in the C43-4 configuration, which corresponds to 0.01pc and $10^3$ au, respectively, for a source distance of 2 kpc. Since an interferometer measures a series of spatial components of an astronomical object through varying projected antenna baselines with respect to the target in an Earth rotation synthesis, one can separate magnetic fields corresponding to different spatial scales through the Stokes $Q$ and $U$ emission, and compare the relative orientation of the multi-scale magnetic fields. 
To probe different structures, we imaged visibilities from the configuration C43-1, and visibilities from the configuration C43-4 after excluding baselines shorter than 250\,m to avoid overlap with the C43-1 data. Such an approach allows a comparison of magnetic fields between independent spatial structures. For clarity, we refer to the data from the C43-4 configuration with the projected baseline restriction to $>$ 250\,m as C43-4-{restricted}.

Fig.~\ref{fig:N6336In} presents magnetic fields in NGC 6334 In produced using data from configurations C43-1, C43-4, and C43-4-restricted, respectively. The position angles of the magnetic field were computed using a 3$\sigma$ cutoff to the Stokes $Q$ and $U$ intensities in their respective images. As shown in Fig.~\ref{fig:N6336In}, the magnetic field orientations measured by the C43-1 and C43-4 configurations (the left and middle panels in Fig.~\ref{fig:N6336In}) are highly correlated despite the difference in angular resolutions. This correlation arises from the fact that the visibilities of the two configurations overlap in their short baselines and sample common structures of the spatially extended emission. Therefore, the magnetic fields of the extended spatial structures mapped by the two array configurations are inherently correlated. However, images made from the C43-4-restricted data reveal only compact spatial structures, as shown in the Stokes $I$ emission in the right panel of Fig.~\ref{fig:N6336In}. For the same reason, the magnetic field information is also associated with these compact structures, which represent the $10^3$~au envelope materials surrounding protostars.

We compare position angles of the magnetic field from the configuration C43-1 and C43-4-restricted, and compute the relative orientation of position angles. Since the magnetic field  is a vector quantity, and there is an ambiguity of 180\deg\ in its direction which is not determined in the linear polarization data, we took the absolute values of the relative orientations and limited the relative angles from $0^\circ$ to $90^\circ$. If pixels in one of the two images do not have magnetic-field measurements, they are excluded in the statistics.

To compute the statistic of ROs of magnetic fields at different spatial scales for the entire sample, we normalize the Stokes $I$, $Q$ and $U$ images to the linear spatial resolution of G14 to account for the difference in source distances. This allows eliminating the bias of the RO statistics to favor closer sources with larger angular sizes. We convovled images to a Gaussian beam of a FWHM angular size equivalent to 3360 au, and 760 au for the two data sets, respectively. We excluded G28 and I18360 from the RO analysis, as there are few independent measurements in the C43-4-restricted images. Therefore, they do not contribute to the overall RO statistic, and including them would require a significantly downgrade in the linear resolution of the data since G28 and I18360 are at larger distances than G14. 

Figure~\ref{fig:angDiff_All} presents the distribution of ROs for the entire sample. Data points from different sources are color coded. It is apparent that ROs of magnetic fields between spatial scales of 0.01~pc and 10$^3$~au are not uniformly distributed over angles ranging from 0\deg\ to 90\deg, and exhibit peaks around 0\deg\ and 90\deg, respectively. The magnetic field data are sampled at 0.5 of the synthesized beam of the C43-4-restricted data; therefore, the number of independent measurements is approximately a factor of 4 lower than the number counts.
A Kolmogorov-Smirnov (K-S) test is performed to determine the probability that this bimodal distribution and randomly distributed angles in a 3-dimensional space are drawn from the same parent population. We find a probability of $6 \times 10^{-6}$ that the two are drawn from the same population. Therefore, such an hypothesis is rejected, and the ROs from the entire sample are not randomly distributed. Magnetic fields at the $10^3$~au envelope scale are correlated with the magnetic fields in the cores. 

\citet{zhang2014} reported a bimodal distribution of ROs by comparing magnetic fields between the parsec-scale clumps and 0.01 to 0.1pc dense cores in a sample of high-mass star-forming regions. The follow-up observations with ALMA in this study reveal a similar bimodal distribution of ROs between the scale of dense cores and envelopes. To further examine RO distributions, we analyze the data for NGC 6334 and G14 separately to avoid the effect of environmental variations in different clouds on the statistics. Both NGC 6334 and G14 clouds were observed in multiple pointings with strong detections of polarized emission that enabled the analysis.

Figure~\ref{fig:angDiff_N6336I} shows the distribution of ROs for NGC~6334 I, In, IV and V. NGC~6334 IR is excluded from this analysis since there is no detection in Stokes $Q$ and $U$ in the C43-4-restricted data.

As shown in Figure~\ref{fig:angDiff_N6336I}, the relative orientation of magnetic fields at a 0.4\arcsec\ resolution with respect to those at a 1\arcsec\ resolution do not appear to be randomly  distributed over angles from 0\deg\ to 90\deg, and exhibits peaks around 0\deg\ and 90\deg, respectively. This bimodal distribution is seen in all four clumps in NGC~6334 I, In, IV and V. Likewise, the magnetic field data are sampled at 0.5 of the synthesized beam of the C43-4-restricted data. A K-S test is performed for each protocluster forming clump to determine the probability that the RO distribution and randomly distributed angles in a 3-dimensional space are drawn from the same parent population. We found a probability for the two distributions drawn from the same population of 0.07, 0.002, $6 \times 10^{-5}$, and 0.4 for NGC~6334 I, In, IV and V, respectively. For NGC~6334 In and IV, the hypothesis is rejected. Magnetic fields at the $10^3$~au envelope scale are either preferentially aligned with or orthogonal to the magnetic field in the cores.

To improve statistics, a relative orientation analysis was also carried out using the data from the four NGC~6334 regions combined, as shown in Figure~\ref{fig:angDiff_N6336All}. A bimodal distribution of ROs remains, with the magnetic field at small scales preferentially aligned with or perpendicular to the field in their parental structures.
A K-S test is performed for NGC 6334 to determine the probability that this distribution and randomly distributed angles in a 3-dimensional space are drawn from the same parent population. We find a probability of $2 \times 10^{-9}$ that the two are drawn from the same population. Therefore, the ROs in NGC 6334 are not randomly distributed, and magnetic fields at $10^3$~au envelope scales are correlated with the magnetic fields in cores.

 In addition to NGC 6334, G14 has significant detections of magnetic fields in multiple pointings, despite the fact that the polarized emission in G14 is fainter than in NGC~6334. While the limited detections of magnetic fields in G14 make the analysis of individual clumps not feasible, we present the combined statistic of all pointings in the G14 cloud in Figure~\ref{fig:angDiff_N6336All}. With an order of magnitude smaller number of data points than in NGC 6334, there appears to be a bimodal distribution of ROs, with a large number of angles near 0\deg\ and 90\deg. We perform a K-S test for G14 to determine the probability that this RO distribution and randomly distributed angles in a 3-dimensional space are drawn from the same parent population. We found a probability of 0.6 that the two are drawn from the same population. Thus, the hypothesis that the RO distribution in G14 is randomly distributed cannot be ruled out due to small number statistics, and it will not be discussed further.

The number of polarization segments detected in G28, G34.41, G35.2N, and I18360 are small in each object. Therefore, we omit the RO analysis for these sources individually. Future observations with sufficient sensitivity will produce additional detections that allow for such analyses.

\section{Discussions}
\label{sec:discussion}

\subsection{RO dependence on densities}

To examine the physical origin of the bimodal distribution of ROs, we analyze ROs of magnetic fields in NGC~6334 using the projected Rayleigh statistic \citep[PRS; see,][and references therein]{jow2018}. 
The PRS tests non-uniformity in a distribution of angles around a particular direction, as quantified by
\begin{equation}
Z_x = \sum^{n}_{i} \frac{\cos(2 \times RO_i)}{\sqrt{n/2}},
\end{equation}
where $n$ is the number of RO angles.
The values $Z_{x}$\,$>$\,0 or $Z_{x}$\,$<$\,0 correspond to ROs clustering around the orientation angles 0\deg\ or 90\deg, respectively.

The PRS was introduced as an extension of the histogram of relative orientation (HRO) technique  \citep{soler2013}. 
In its initial formulation, the HRO is based on an RO histogram.
The clustering around preferential orientations 0\deg\ or 90\deg\ was calculated using the histogram counts in the range from 0\deg\ to 22\deg.5 and from 67\deg.5 to 90\deg, respectively.
The PRS circumvents the histogram construction and provides a direct estimate of the angle clustering significance \citep[see,][for a comparison between $Z$ and the relative orientation parameter, $\xi$]{soler2019}.
To enable the comparison of sources with different numbers of polarization detections, the PRS can be mapped into the range $-1$ to $1$ through normalization by its theoretical maximum value \citep[see, for example,][]{mininni2025}, that is,
\begin{equation}
\frac{Z_x}{Z_{x,{\rm max}}} = \sum^{n}_{i}\cos(2 \times RO_i),
\end{equation}
which is also referred to as the alignment measure \citep[AM; e.g.,][]{lazarian2018}.

Figure~\ref{fig:N6334all_PRS} presents the normalized PRS as a function of column densities for the NGC 6334 observations shown in Figure~\ref{fig:angDiff_N6336All}. We used the continuum image from the combined C43-1 and C43-4 configuration to derive column densities. The free-free emission from the ionized gas associated with UCH{\sc ii} regions has been detected in NGC 6334 I and V. However, their contributions are negligible compared to the flux detected in the 1.3-mm continuum emission from the ALMA band 6. Thus, the continuum emission is mainly attributed to dust gains \citep[e.g.,][]{liu_J2023a}. We converted the 1.3-mm continuum emission to column densities, using the dust opacity law $\kappa =  ({\nu \over THz})^\beta$ \citep{hildebrand1983} and a dust emissivity spectral index $\beta$ of 1.5~\citep{beuther2007b}. This formulation results in a dust opacity of 1 $cm^2~g^{-1}$ at 230 GHz, which is consistent with the opacity for dust grains with ice mantles at gas densities between 10$^6$ and 10$^8$ $cm^{-3}$~\citep{ossenkopf1994}. The dust temperature is approximated by the gas temperature reported in \citet{liu_J2023a}, who derived the temperature distribution in the region using four CH$_3$OH transitions with upper energy levels between 45 to 508\,K. The derived gas temperatures are typically 100\,K, and reach more than 300\,K in some localized cores. 

For comparison, we also derive ROs between magnetic field orientations and plane-of-the-sky gravitational forces \citep[see also ][]{liu_J2023a},
which are computed using the column density data as follows.
\begin{equation}
\vec{g}_{ij}=\sum^{i\neq i',j\neq j'}_{i'j'}G\frac{(N_{\rm
H})_{ij}(N_{\rm H})_{i'j'}}{\left\lVert
\vec{r}_{ij}-\vec{r}_{i'j'}\right\rVert^{3} }(\vec{r}_{ij}-\vec{r}_{i'j'}),
\end{equation}
where the indices $i$, $j$ and $i'$, $j'$ correspond to pixels in each
column density map \citep{liu_J2023a}.
Figure~\ref{fig:N6334all_PRS_LG} presents the ROs between local gravity and magnetic fields as a function of column densities. It is apparent that within the density range from $5 \times 10^{22}$ to $2 \times 10^{24}$ $cm^{-2}$, the magnetic field is mostly parallel to the direction of the inferred gravitational force.

It is also apparent in Figure~\ref{fig:N6334all_PRS} that at column densities greater than $10^{23}$\,cm$^{-2}$, the magnetic field is mostly parallel to the $N_{{\rm H}_{2}}$ structures and the direction of local gravity. In this density range, the magnetic field is also preferentially parallel to the field in the parent dense core.
Previous analysis of ROs between magnetic fields and density contours on scales between tens of parsecs and down to 0.1\,pc found a global trend in star-forming regions in the solar neighborhood: The ROs progressively change from preferentially parallel to preferentially perpendicular at column densities around $N_{\rm H}$\,$\approx$\,$10^{21.7}$\citep{planck2016b,soler2019}.
This pattern has been interpreted as the result of the transition from sub- or trans-Alfv\'{e}nic to super-Alfv\'enic turbulence with increasing densities \citep{chen2016ApJ_829_84C} and, more recently, as the sequence of the combined effect of anisotropy, caused by a relatively strong magnetic field, and a converging flow, such as that produced by gravitational collapse \citep{soler2017b,seifried2020b,ibanez-mejia2022}.
Our results suggest that this trend is not extended to smaller scales and higher column densities.

Observations of magnetic field orientations covering scales below 0.1\,pc indicate that the ROs at $N_{\rm H}$\,$>$\,$10^{21.7}$ are not indefinitely perpendicular but tend to return to a parallel configuration at higher column densities \citep{pillai2020, koch2022}.
Numerical simulations covering spatial scales between 80,000\,AU ($\sim$\,0.4\,pc) and 100\,AU indicate that even large-scale magnetic fields cannot maintain a unique trend across this broad range of scales \citep[see, for example,][]{hull2017b}.
The analytic study of the relative orientation between the density structures and the magnetic fields presented in \citet{soler2017b} indicates that preferential orientations parallel or perpendicular constitute equilibrium points of the MHD equations. 
Thus, these ROs tend to be more represented in the ISM. 
\cite{soler2017b} and \cite{seifried2020b} indicate that the transition between the two configurations is driven by the increase in the velocity divergence, which can be induced by gravitational collapse or shocks.
\cite{ibanez-mejia2022} shows that the transition between parallel and perpendicular observed in \citet{planck2016b} can be explained by the change between regions dominated by the Lorentz force, at low densities, and by the gravitational force, at higher densities.

The trend in Fig.~\ref{fig:N6334all_PRS} corresponds to spatial scales and density regimes in which gravitational forces dominate over magnetic forces. The evidence supporting this argument is as follows. First, as shown in the right panel of Fig.~\ref{fig:N6336In}, only compact emission remains. In addition, as shown in Fig.~\ref{fig:N6334all_PRS_LG}, the magnetic field tends to be aligned with the direction of the gravitational force. Furthermore, \citet{liu_J2023a} analyzed the JCMT and ALMA full polarization observations of NGC~6334 using the polarization-intensity gradient method \citep{koch2012a}. They found that the mass-to-magnetic-flux ratio derived from the comparison between the magnetic force and the gravitational force increases with column densities. A transition density from a subcritical state to a supercritical state is at $3 \times 10^{22}$ \cmm.
The preferentially parallel ROs indicate that the magnetic field follows the density structures driven by gravitational collapse, as in cases where the field follows the streamer features in the polarization observations reported in \cite{beuther2024,sanhueza2021,koch2022,pan2024}.
Following the argument in \cite{soler2017b}, the preferentially parallel RO suggests an additional change in velocity divergence at scales below 0.1\,pc, which can be assigned to gravitational collapse at an envelope scale within dense cores.
The result in Fig.~\ref{fig:N6334all_PRS} expands previous RO studies to higher densities and smaller spatial scales, covering a regime where ROs are yet to be systematically studied in numerical simulations.

ROs of magnetic fields at different spatial scales exhibit a high likelihood of parallel or perpendicular alignments. This behavior is found in ROs between magnetic fields in parsec-scale clumps and $0.01 - 0.1$~pc dense cores within \citep{zhang2014}. This distribution of ROs may suggest that these clumps are initially in a sub-Alfv\'enic state, which is supported by observations of the clumps in NGC~6334 \citep{li_H2015}. As gravitational collapses ensue within the clumps to form dense molecular cores, the gas transition from a sub-Alfv\'enic to super-Alfv\'enic state as the gas densities increase, and ROs change from parallel to perpendicular alignments. As dense cores continue to collapse and fragment, the ROs of magnetic fields between the envelope scales and their parental cores show a preferential parallel alignment. This change may represent another transition as suggested in other studies \citep{pillai2020} at even higher densities as gravitational forces become dominant over the magnetic forces. Therefore, the bimodal distribution in ROs between multi-scale magnetic fields observed in \citet{zhang2014} and here may reflect a change in magnetic field alignments in different density regimes. Further numerical studies of protostellar cluster forming clumps exploring spatial scales smaller than a few 100 au are needed to shed more light on the underlying physics that gives rise to the change of magnetic field alignments.

\section{Summary}
\label{sec:sum}
We present ALMA full polarization observations for 17 protostellar cluster forming clumps previously studied with the SMA in \citet{zhang2014}. The linearly polarized continuum emission at 1.4~mm revealed the plane-of-the-sky component of magnetic fields at a resolution up to 0\arcsec.4, corresponding to the $10^3$ au envelope scale. The data from both C43-1 and C43-4 configurations allowed multi-scale analysis of magnetic fields. We compare magnetic field orientations from C43-1 with those from C43-4, but excluding visibilities with UV distances overlapping with the C43-1 data. The main findings are summarized as follows.

\begin{itemize}
     
    \item Magnetic fields are detected in nearly in all clumps ranging from infrared dark clouds to clumps harboring high-mass protostellar objects.
    
    \item The relative orientations of magnetic fields between the configuration C43-1 data ($\sim$ 1\arcsec\ resolution) and the C43-4-restricted data that excludes visibilities of baselines shorter than 250 m exhibit a bimodal distribution with peaks near 0\deg\ and 90\deg\ for the entire sample as well as for NGC~6334, which indicates that the magnetic field at the $10^3$~au envelope scale tends to be either aligned with or orthogonal to the magnetic field in 0.01~pc dense cores. A projected Rayleigh statistic analysis of ROs found an overall preferential parallel alignment between the two magnetic fields.

    \item This RO distribution likely arises from gravitational collapse in envelope scales as the magnetic field is dragged by the infalling motions in the magnetically supercritical gas.

\end{itemize}

\begin{acknowledgments}

The authors thank the anonymous referee for constructive comments that helped to improve the paper.
JMG acknowledges support by the grant PID2020-117710GB-I00 and
PID2023-146675NB-I00 (MCI-AEI- FEDER, UE), and partial support by the program Unidad de Excelencia Mar{\'i}a de Maeztu CEX2020-001058-M. ZYL is supported in part by NASA 80NSSC20K0533, NSF AST-2307199, and the Virginia Institute of Theoretical Astronomy (VITA). This paper makes use of the following ALMA data: ADS/JAO.ALMA$\#$2017.1.00793.S. ALMA is a partnership of ESO (representing its member states), NSF (USA) and NINS (Japan), together with NRC (Canada), MOST and ASIAA (Taiwan), and KASI (Republic of Korea), in cooperation with the Republic of Chile. The Joint ALMA Observatory is operated by ESO, AUI/NRAO and NAOJ. 

\end{acknowledgments}
\clearpage

\begin{figure}[!h]
\figurenum{1a}
\centering 
\includegraphics[width=0.49\textwidth]{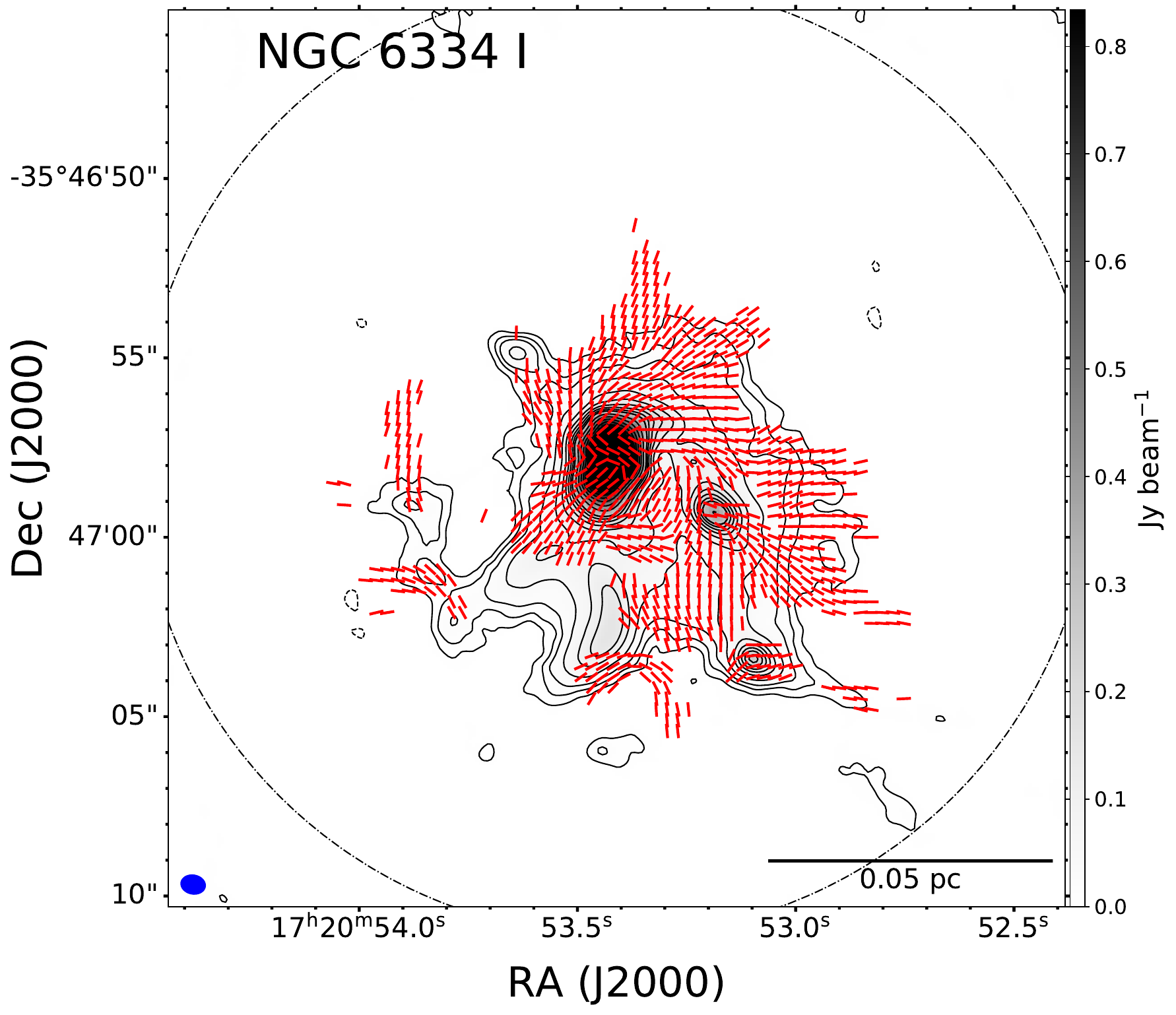}
\includegraphics[width=0.49\textwidth]{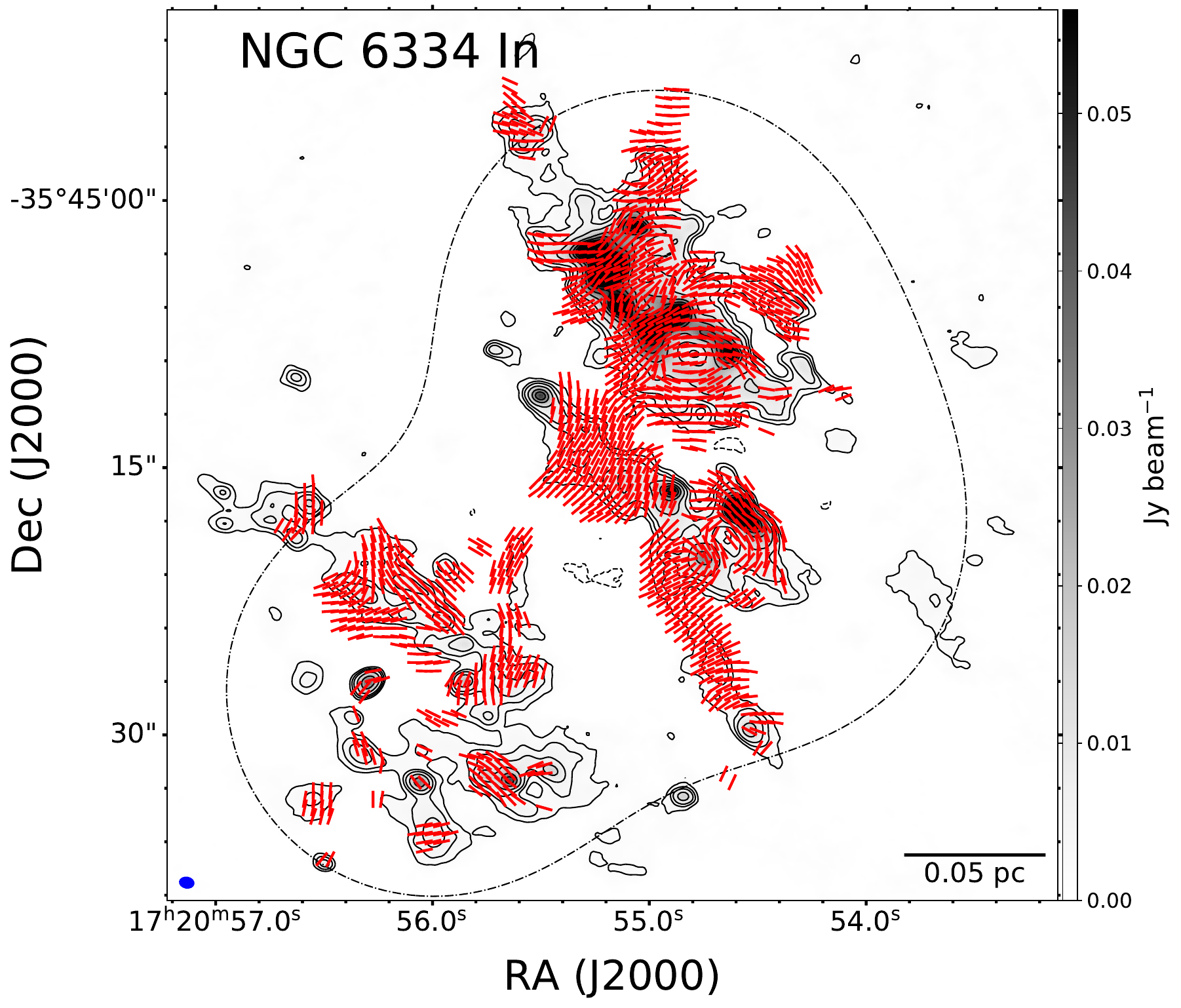}
\includegraphics[width=0.49\textwidth]{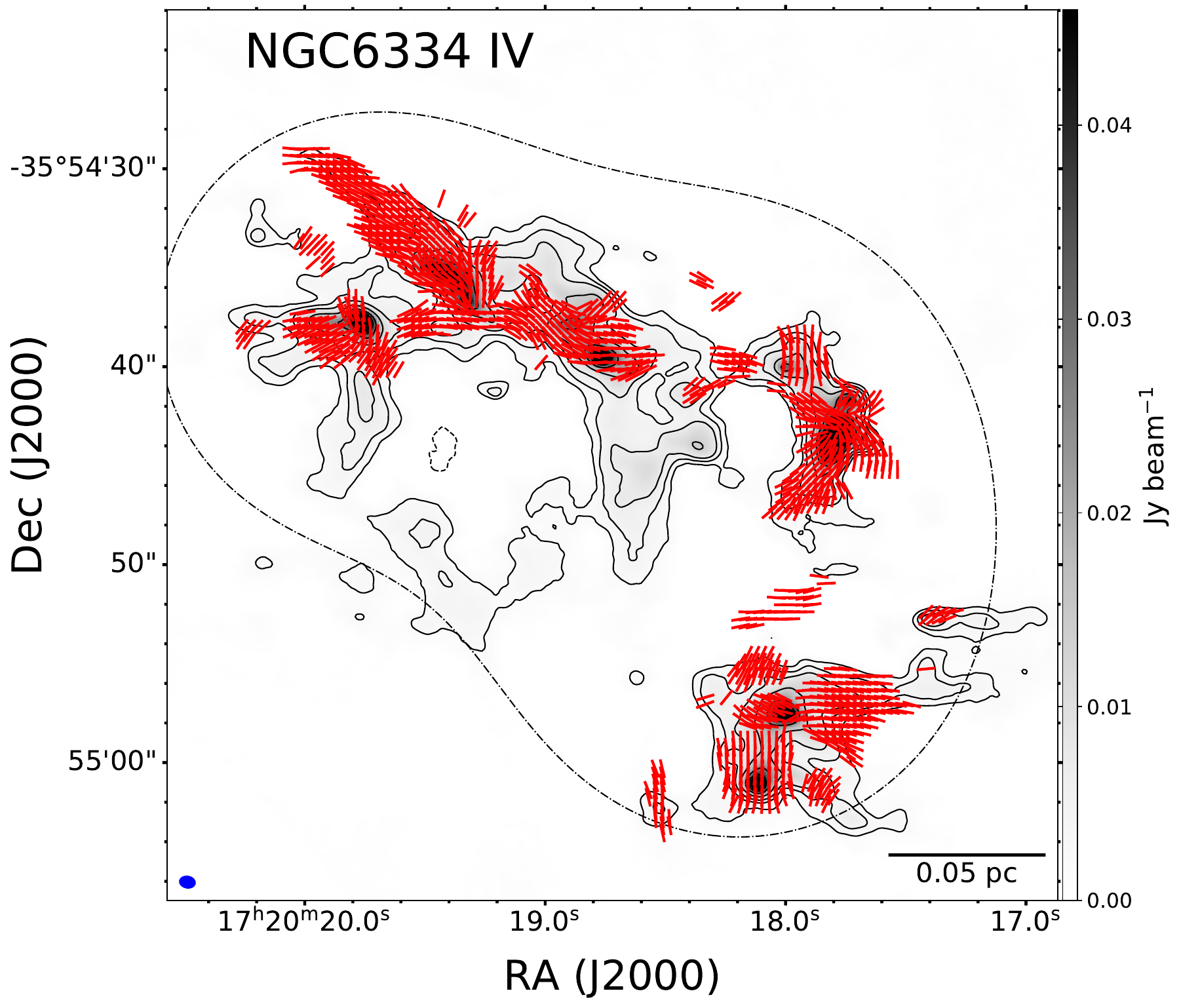}
\includegraphics[width=0.49\textwidth]{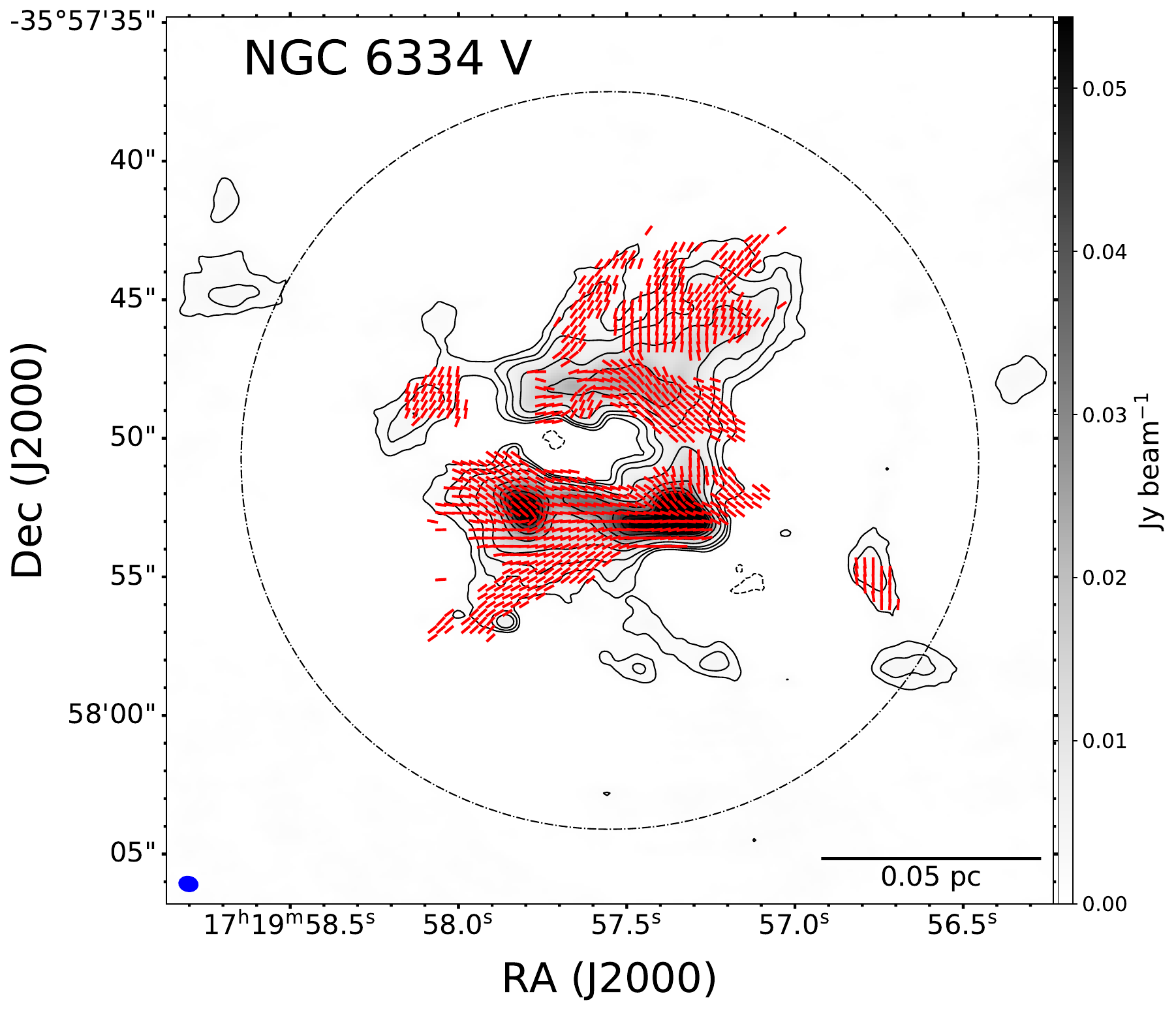}
\caption{Continuum emission at 225\,GHz and magnetic fields for NGC 6334 I, In, IV and V. The Stokes $I$ emission is shown in contours and in grey scales. The red line segments represent the position angle (PA) of the magnetic field plane-of-the-sky component, produced using a cutoff threshold of 3$\sigma$ in Stokes $Q$ and $U$ intensities. The synthesized beam is marked at the lower-left corner of each panel. The horizontal bar at the lower-right corner of each panel represent a linear scale of 0.05 pc. The dotted line marks the 50\% sensitivity level of the observations. Unless stated otherwise, all images in Fig. 1 were made from combined visibilities from the configuration C43-1 and C43-4. The Stokes $I$ intensities are contoured at ($\pm 3$, $\pm 6$, 10, 20, 40, 60, 100) $\times  \sigma$, where $\sigma$ values are listed in Table~\ref{tab:2}.}
\label{fig1:N6334}
\end{figure}

\begin{figure}[!h]
\figurenum{1b}
\centering 
\includegraphics[width=0.49\textwidth]{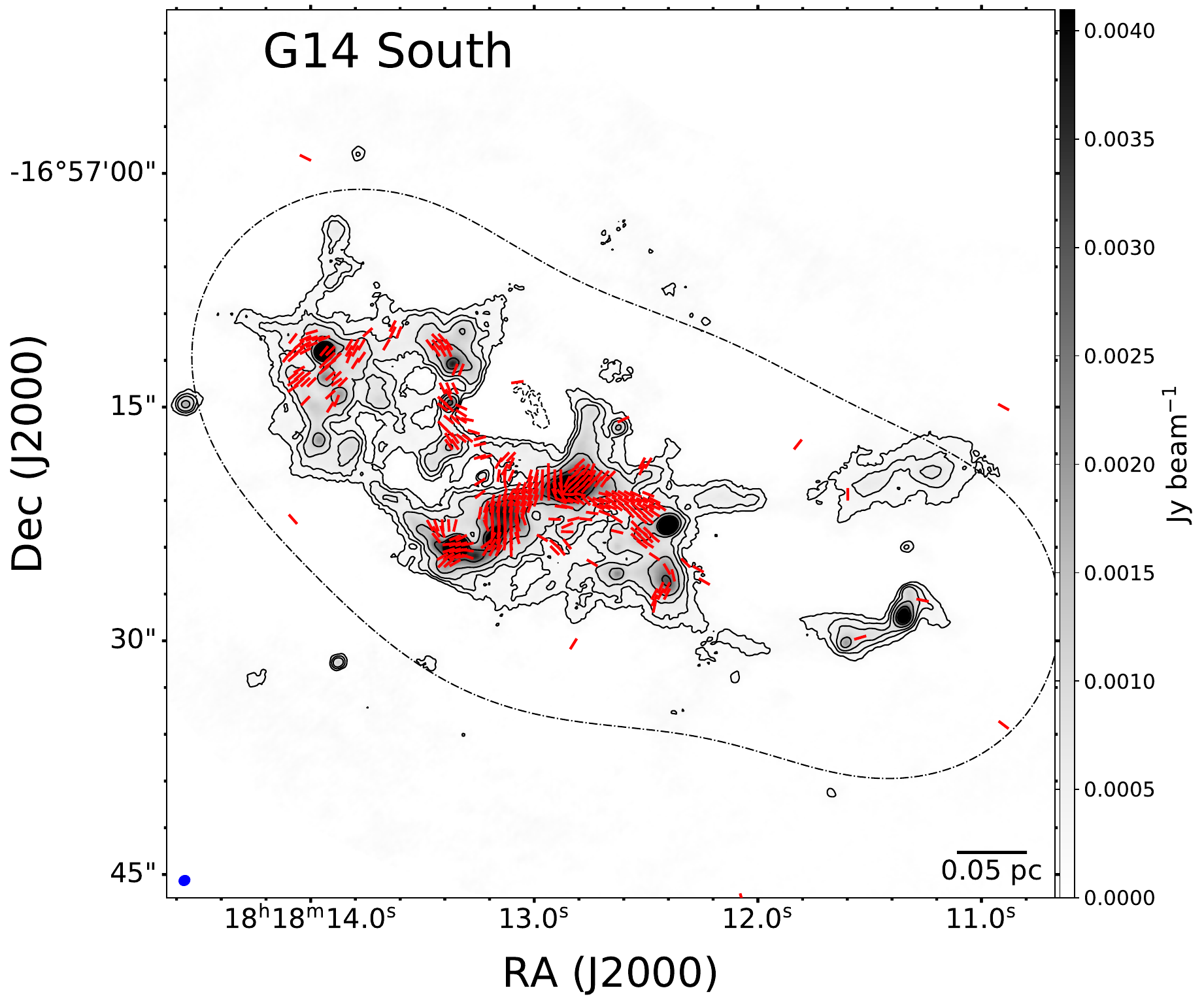}
\includegraphics[width=0.49\textwidth]{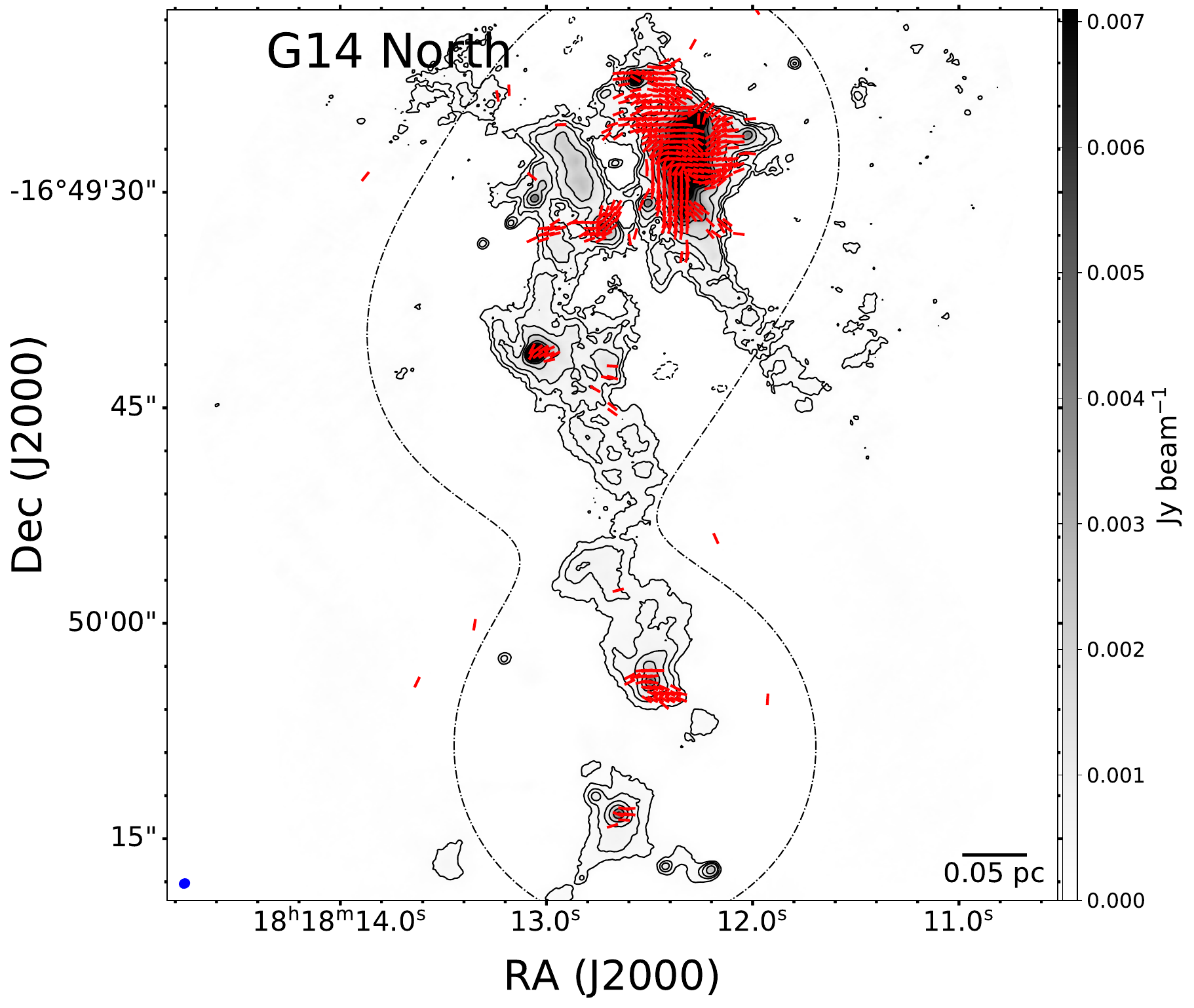}
\caption{Continuum emission at 225 GHz and magnetic fields for G14. The Stokes $I$ emission is shown in contours and grey scales. The red line segments represent the position angles (PA) of the plane-of-the-sky component of magnetic fields. The synthesized beam is marked at the lower-left corner of each panel. The horizontal bar at the lower-right corner of each panel represent a linear scale of 0.05 pc. The dotted line marks the 50\% sensitivity level of the observations.}
\label{fig1:G14}
\end{figure}

\begin{figure}[!h]
\figurenum{1c}
\centering 
\includegraphics[width=0.49\textwidth]{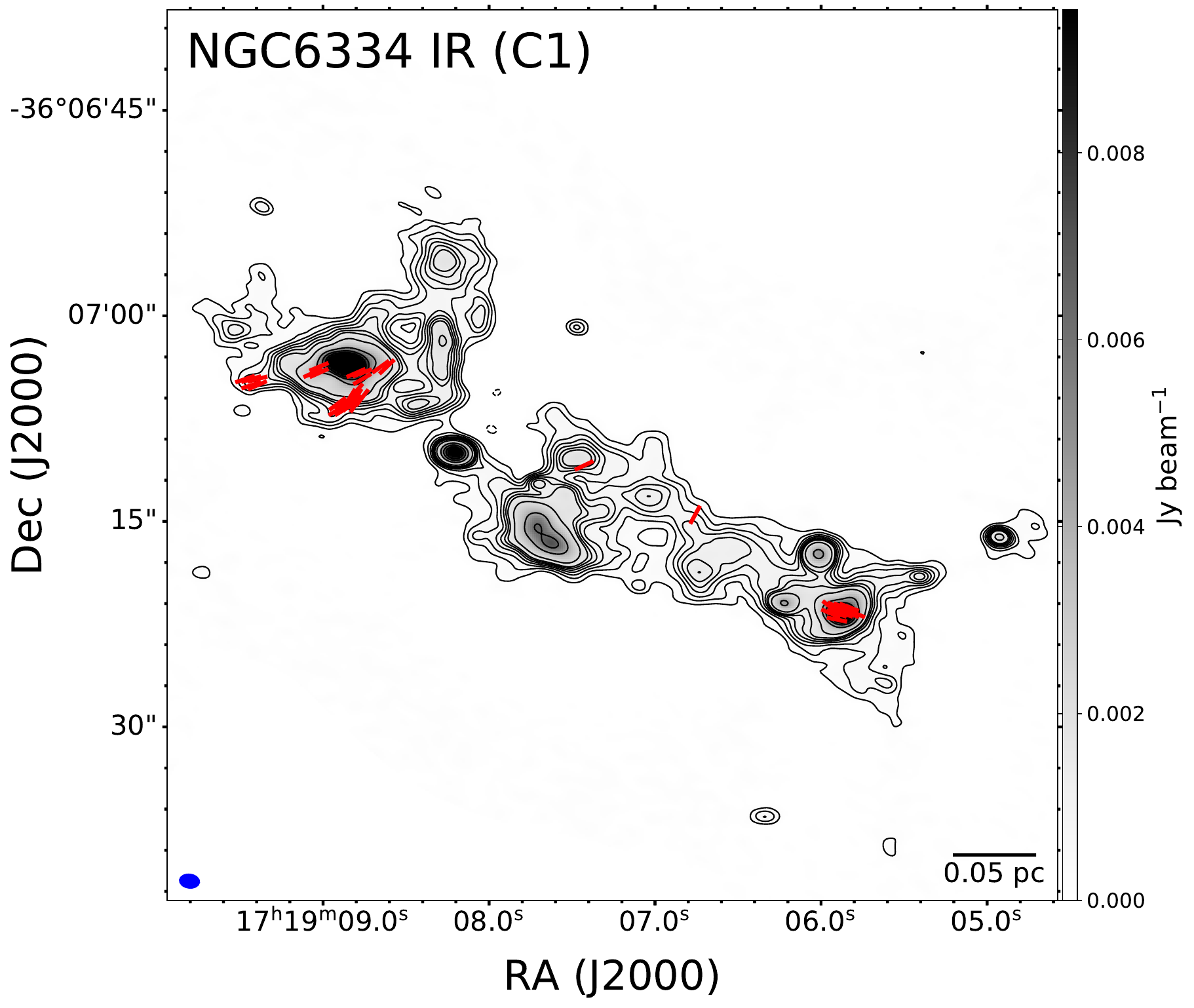}
\includegraphics[width=0.49\textwidth]{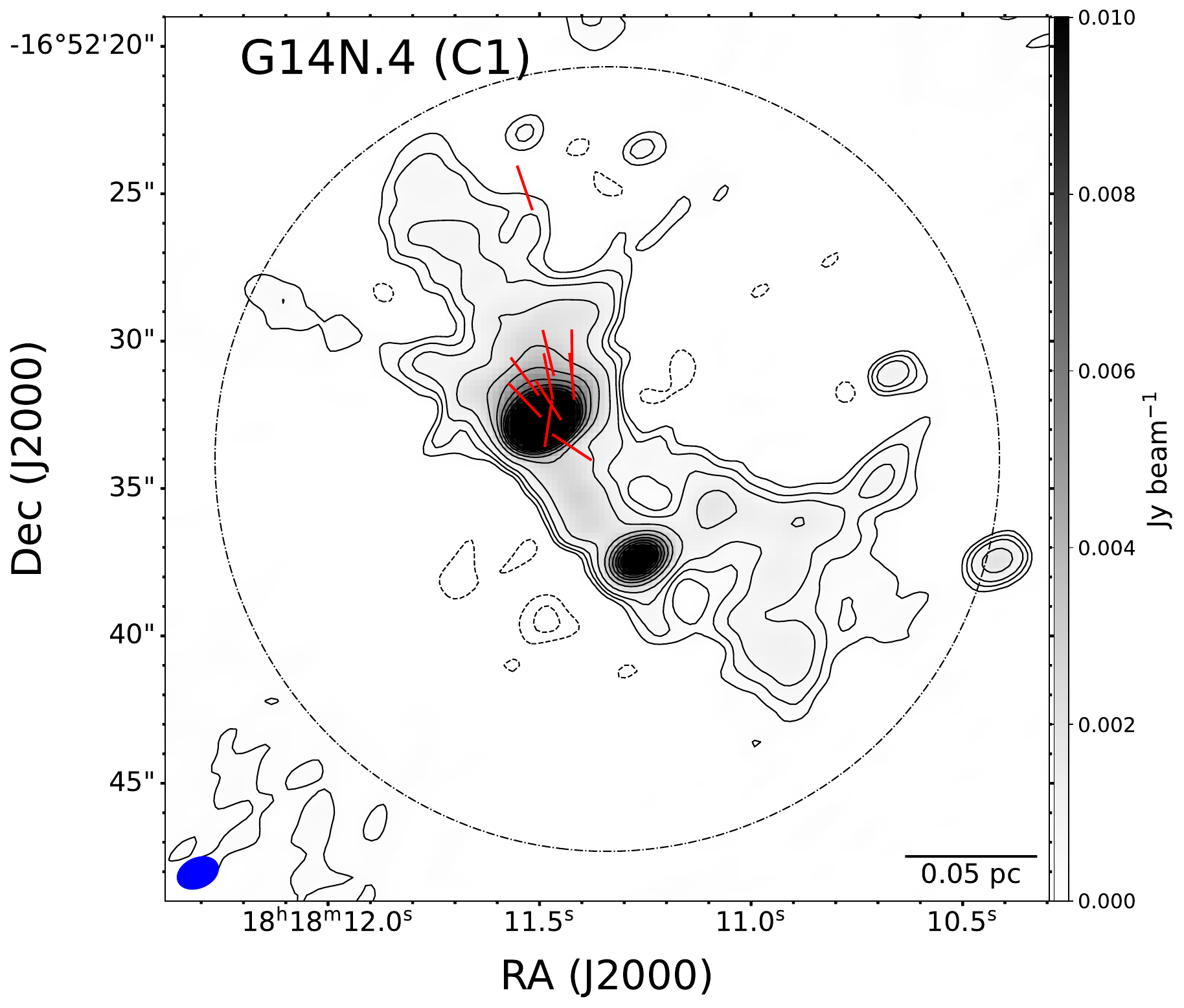}
\includegraphics[width=0.49\textwidth]{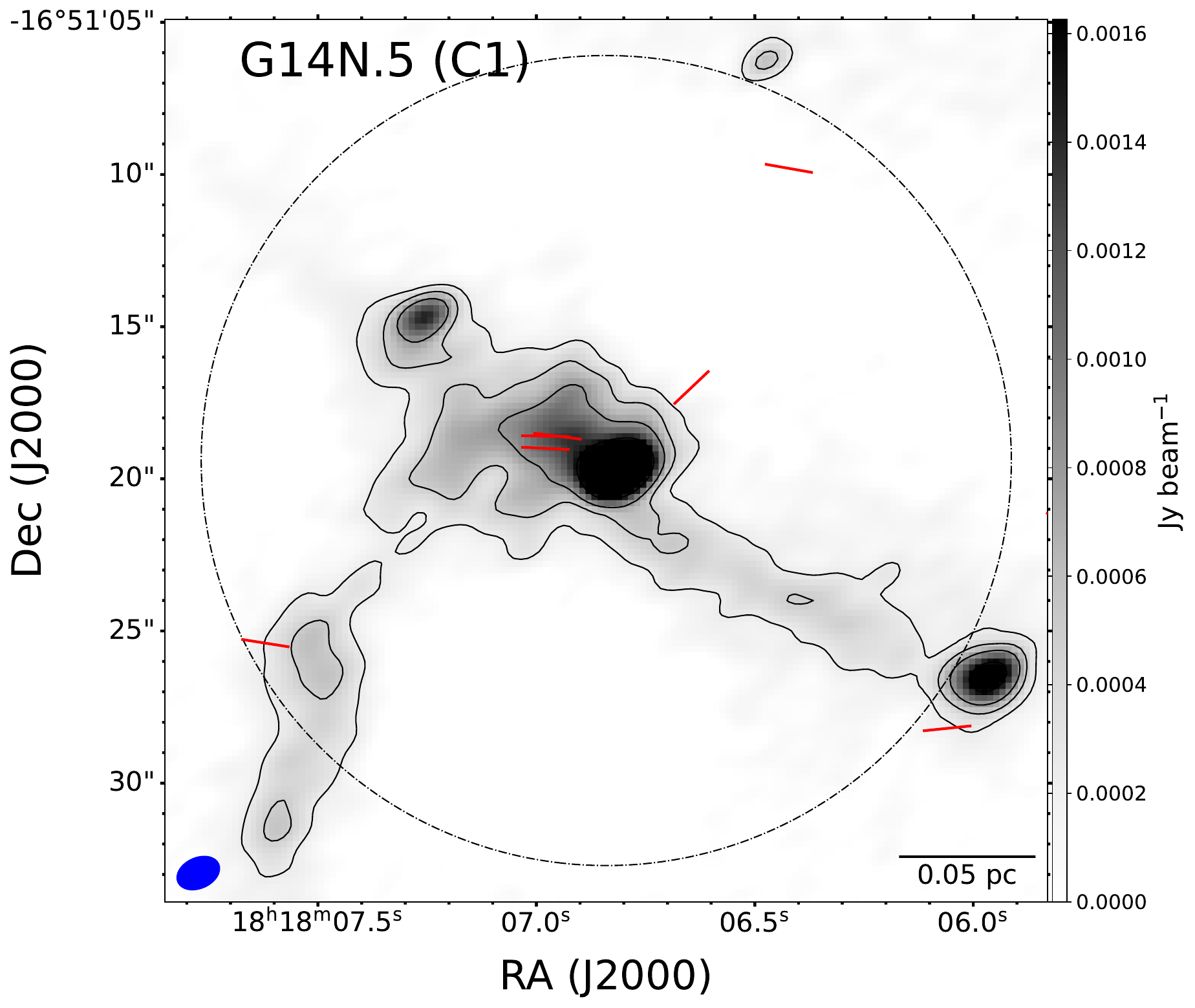}
\includegraphics[width=0.49\textwidth]{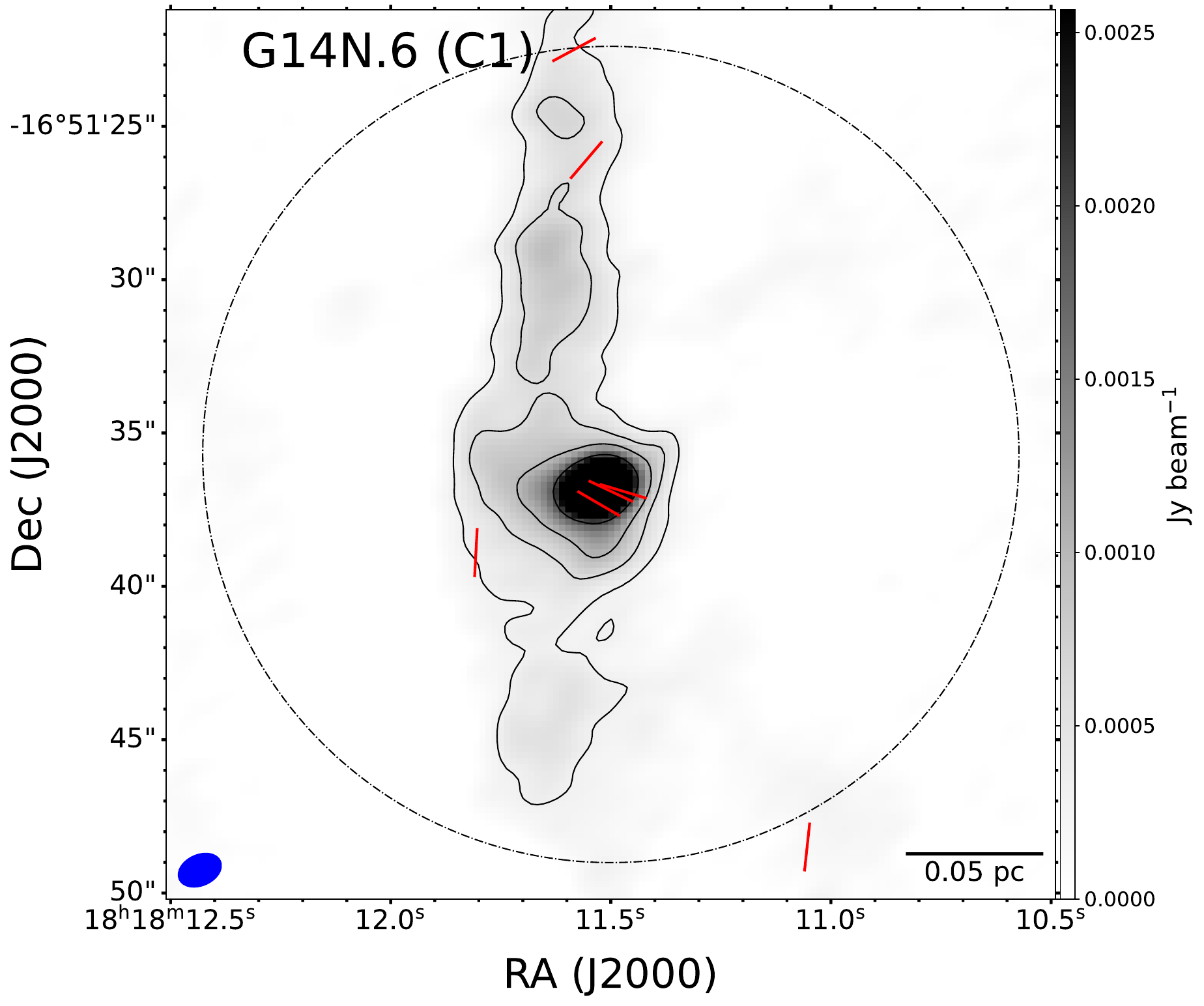}
\caption{Continuum emission at 225 GHz for NGC 6334IR, G14N.4, G14N.5 and G14N.6. The Stokes $I$ emission is shown in contours and in grey scales. The red line segments represent the position angles (PA) of the plane-of-the-sky component of magnetic fields. The synthesized beam is marked at the lower-left corner of each panel. The horizontal bar at the lower-right corner of each panel represent a linear scale of 0.05 pc. The dotted line marks the 50\% sensitivity level of the observations. Images were made from visibilities from the configuration C43-1 only. }
\label{fig1:G14_5-7}
\end{figure}

\begin{figure}[!h]
\figurenum{1d}
\centering 
\includegraphics[width=0.49\textwidth]{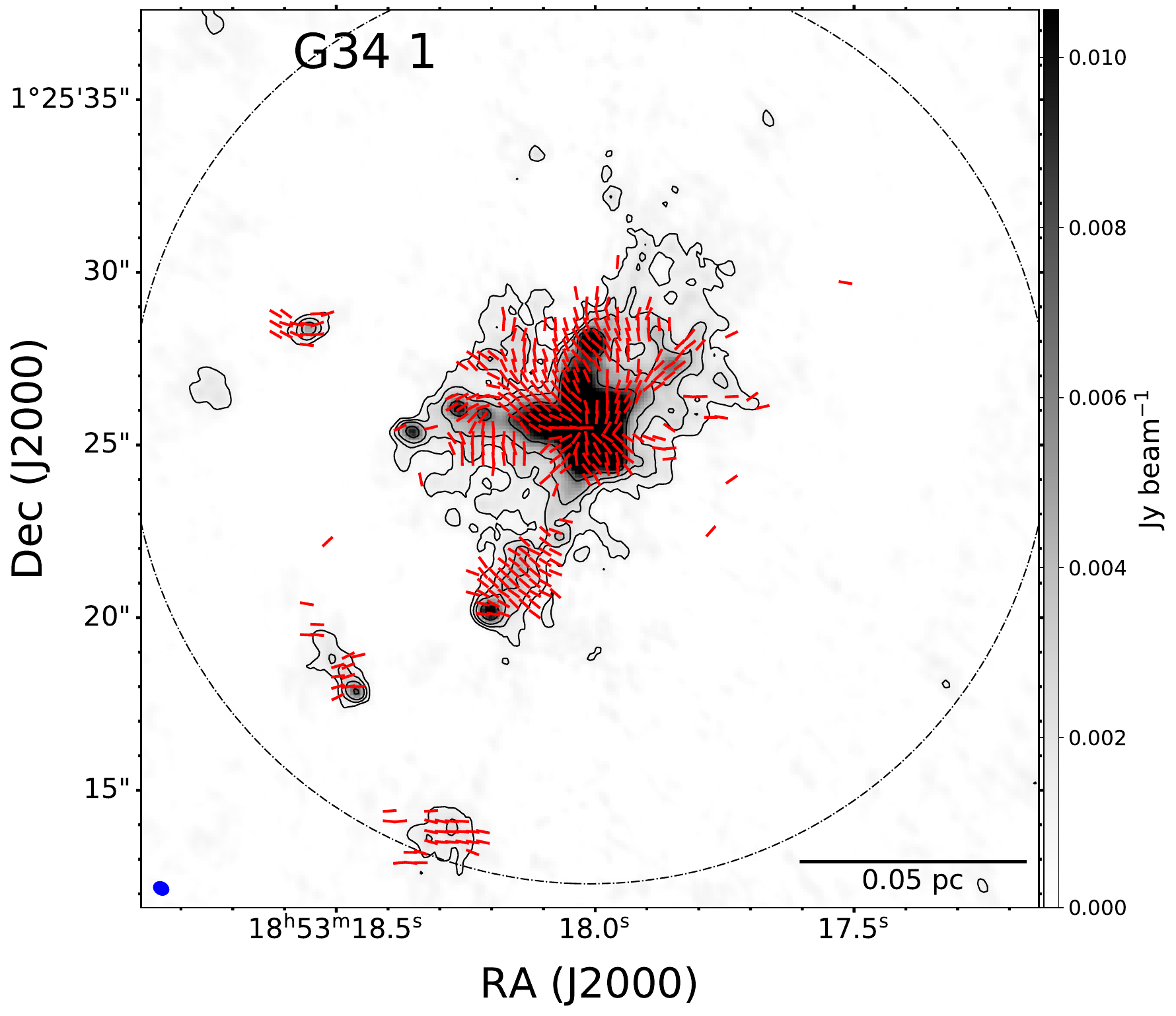}
\includegraphics[width=0.49\textwidth]{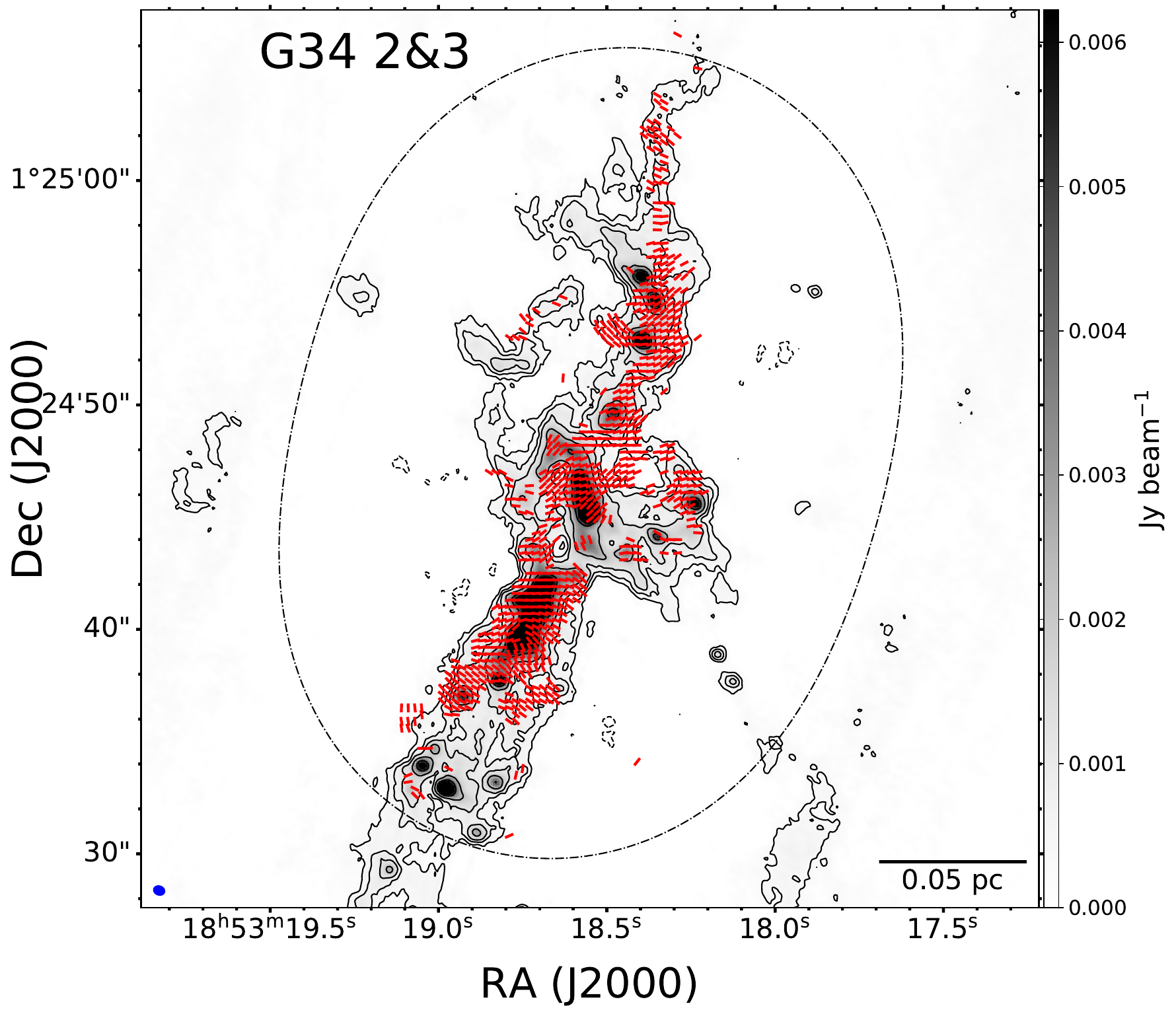}
\caption{Continuum emission at 225 GHz and magnetic fields for G34. The Stokes $I$ emission is shown in contours and grey scales. The red line segments represent the position angles (PA) of the plane-of-the-sky component of magnetic fields. The synthesized beam is marked at the lower-left corner of each panel. The horizontal bar at the lower-right corner of each panel represent a linear scale of 0.05 pc. The dotted line marks the 50\% sensitivity level of the coverage.}
\label{fig1:G34}
\end{figure}

\begin{figure}[!h]
\figurenum{1e}
\centering 
\includegraphics[width=0.49\textwidth]{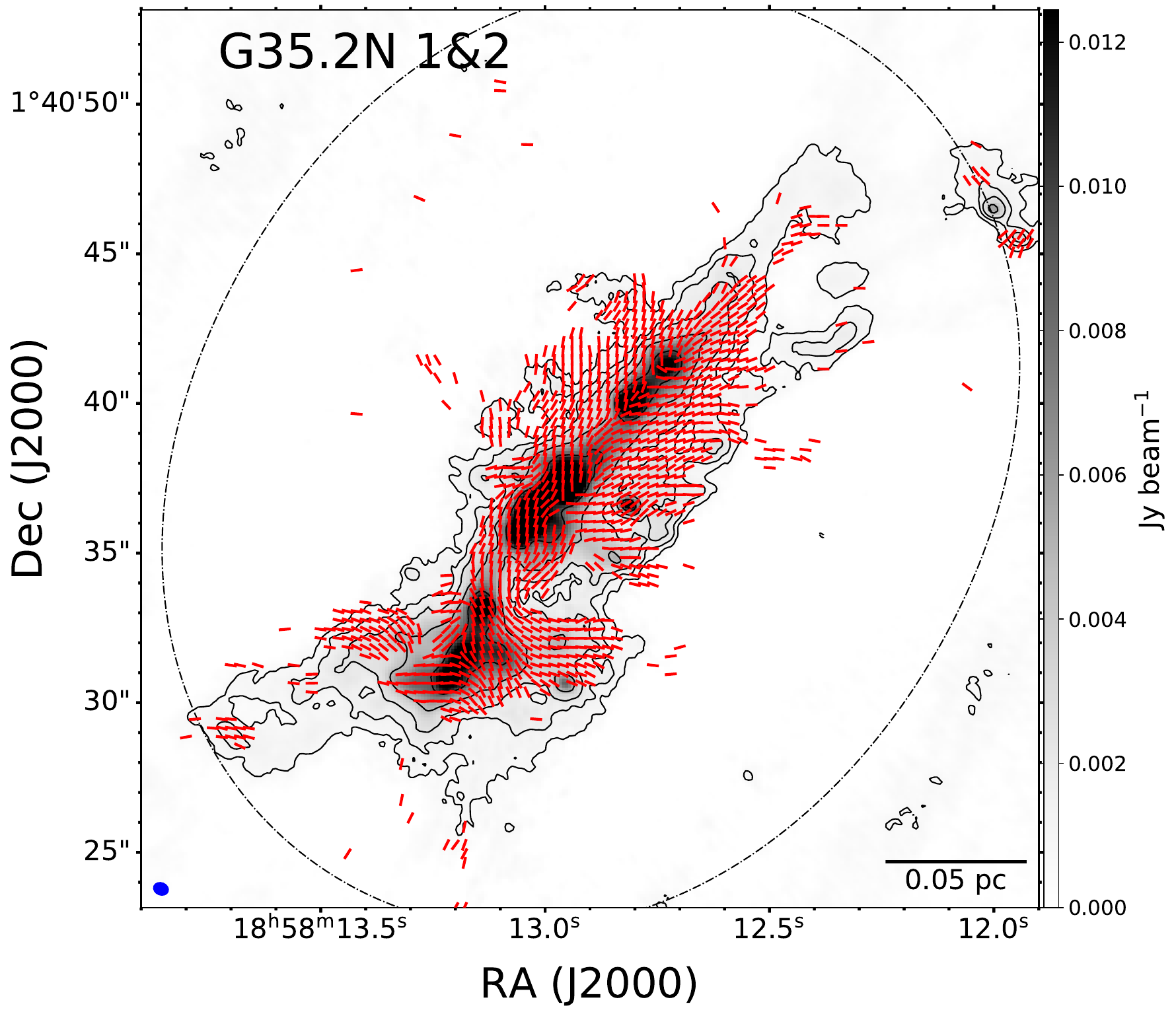}
\includegraphics[width=0.49\textwidth]{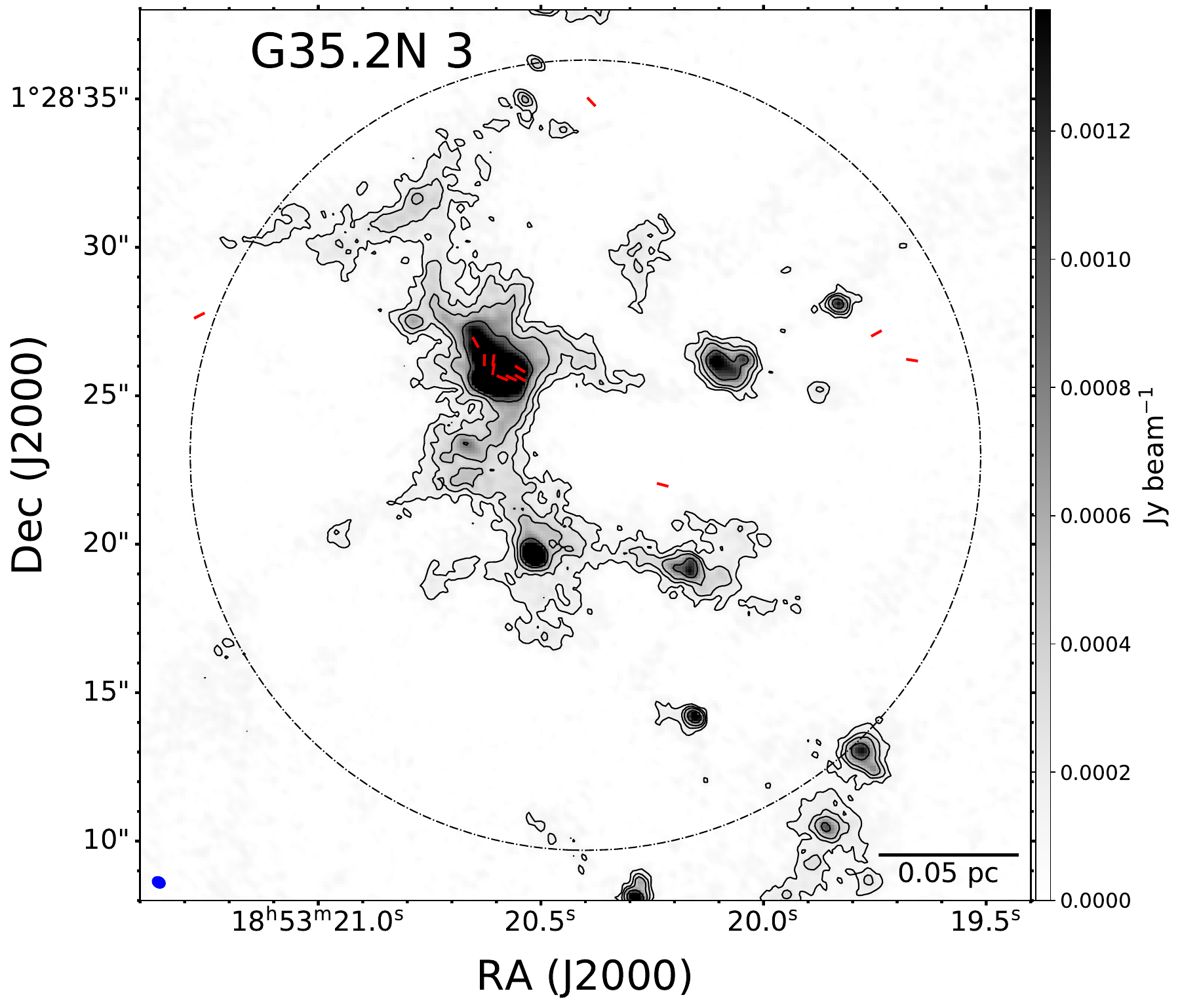}
\caption{Continuum emission at 225 GHz and magnetic fields for G35.2N. The Stokes $I$ emission is shown in contours and grey scales. The red line segments represent the position angles (PA) of the plane-of-the-sky component of magnetic fields. The synthesized beam is marked at the lower-left corner of each panel. The horizontal bar at the lower-right corner of each panel represent a linear scale of 0.05 pc. The dotted line marks the 50\% sensitivity level of the observations.}
\label{fig1:G35}
\end{figure}

\begin{figure}[!h]
\figurenum{1f}
\centering 
\includegraphics[width=0.32\textwidth]{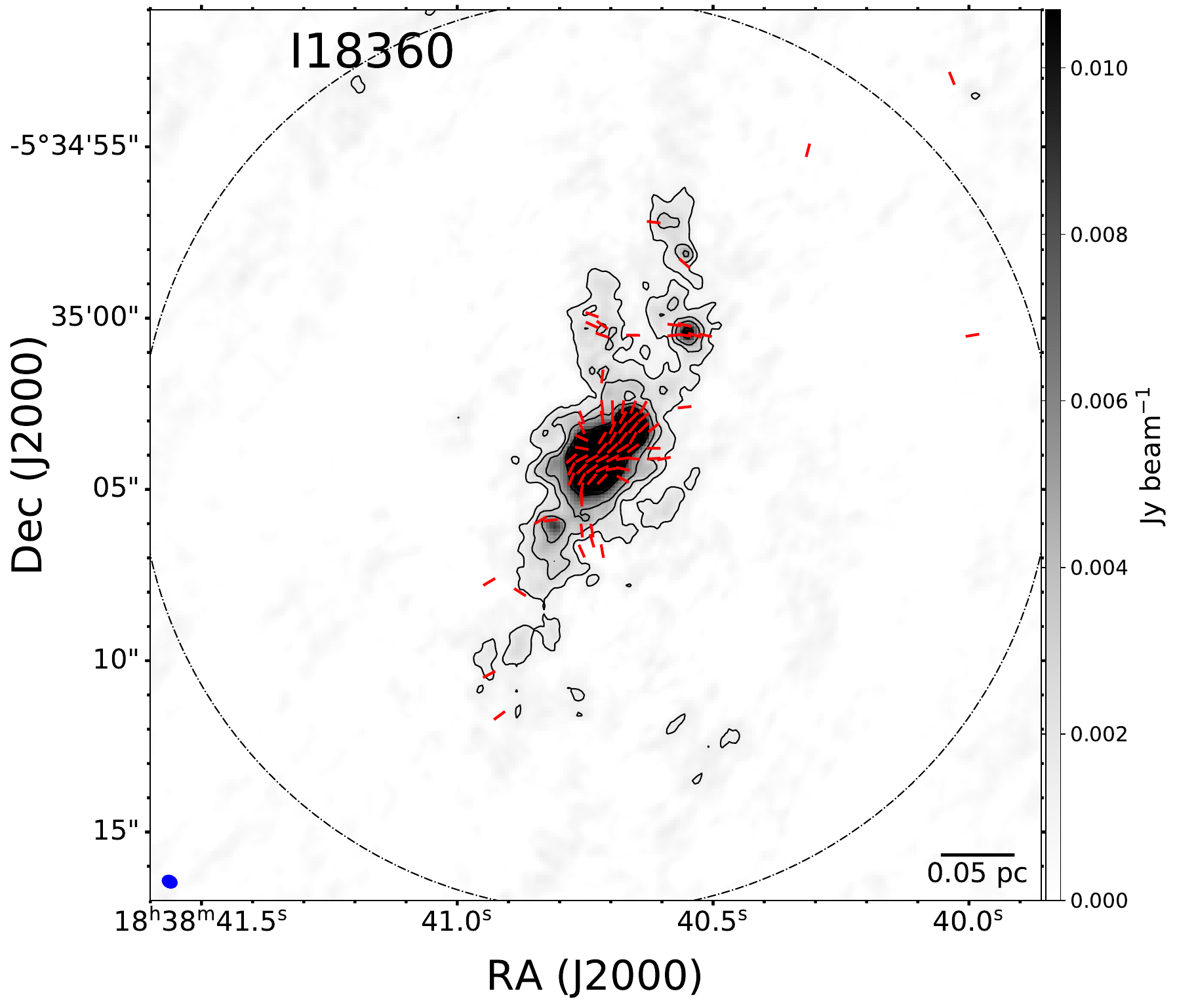}
\includegraphics[width=0.32\textwidth]{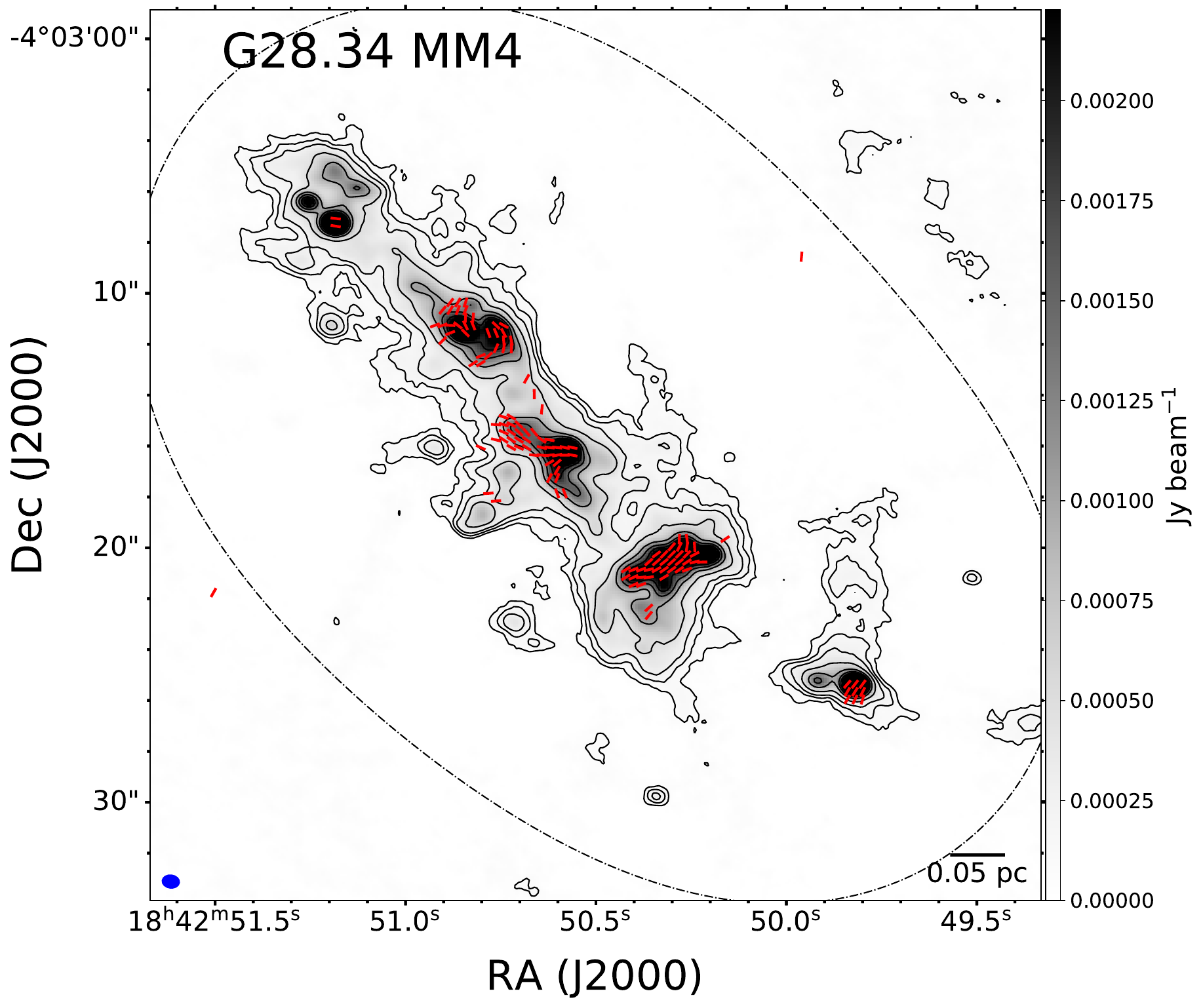}
\includegraphics[width=0.32\textwidth]{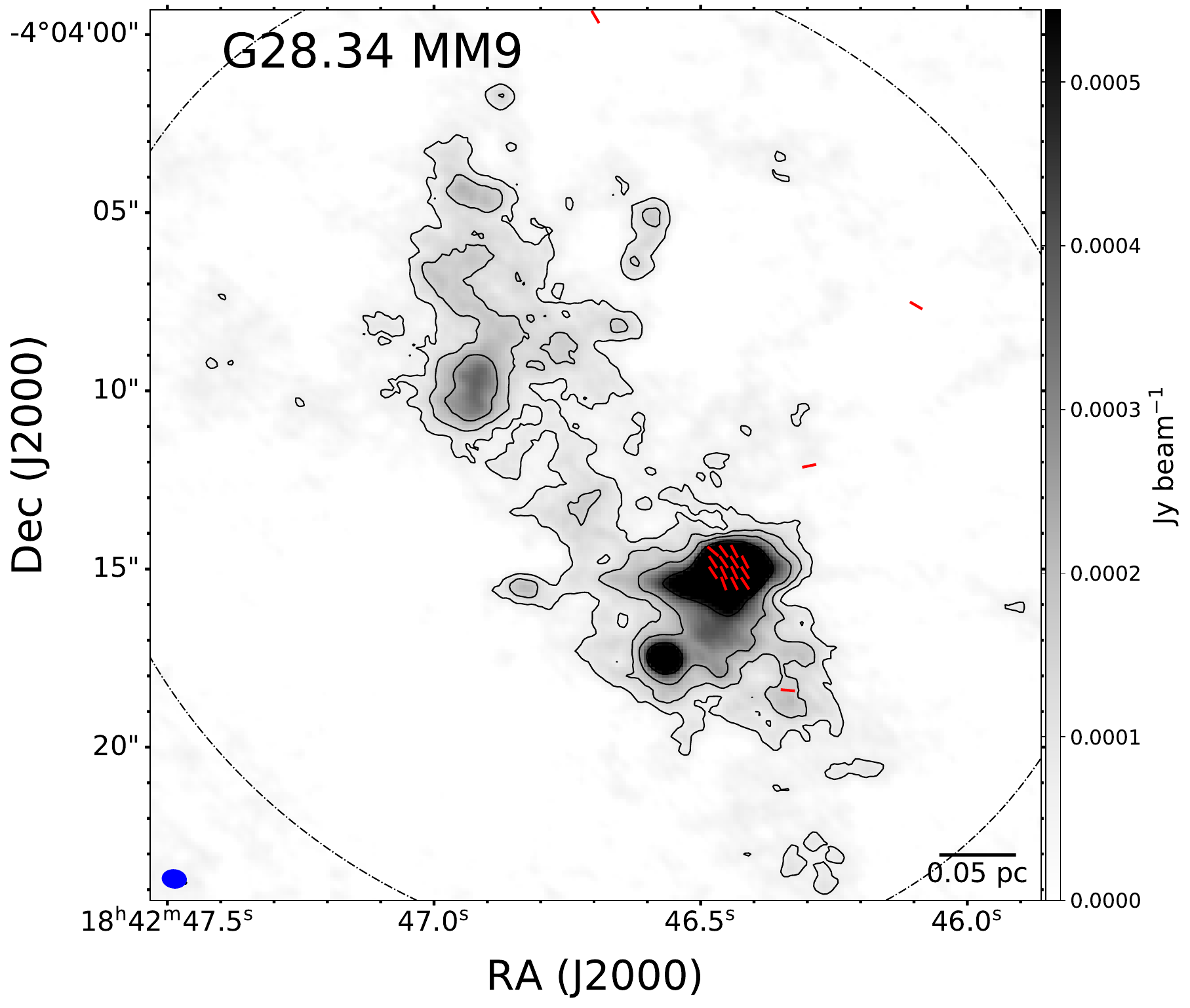}
\caption{Continuum emission at 225 GHz and magnetic fields for IRAS 18360 and IRDC G28.34. The Stokes $I$ emission is shown in contours and grey scales. The red line segments represent the position angles (PA) of the plane-of-the-sky component of magnetic fields. The synthesized beam is marked at the lower-left corner of each panel. The horizontal bar at the lower-right corner of each panel represent a linear scale of 0.05 pc. The dotted line marks the 50\% sensitivity level of the observations.}
\label{fig1:I18360_G28}
\end{figure}

\setcounter{figure}{2}
\begin{figure}[!ht]
\figurenum{2}
\centering 
\includegraphics[width=0.32\textwidth]{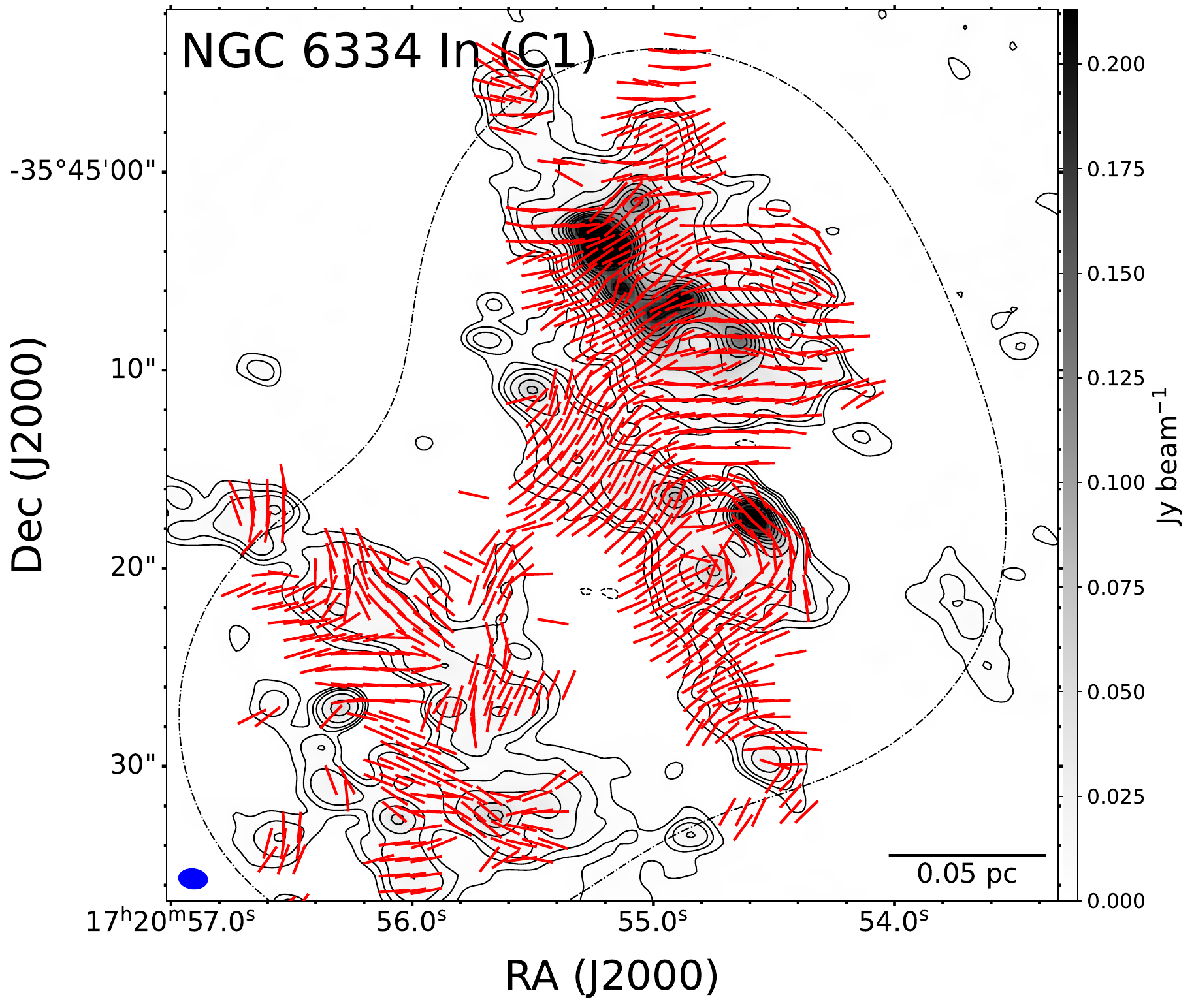}
\includegraphics[width=0.32\textwidth]{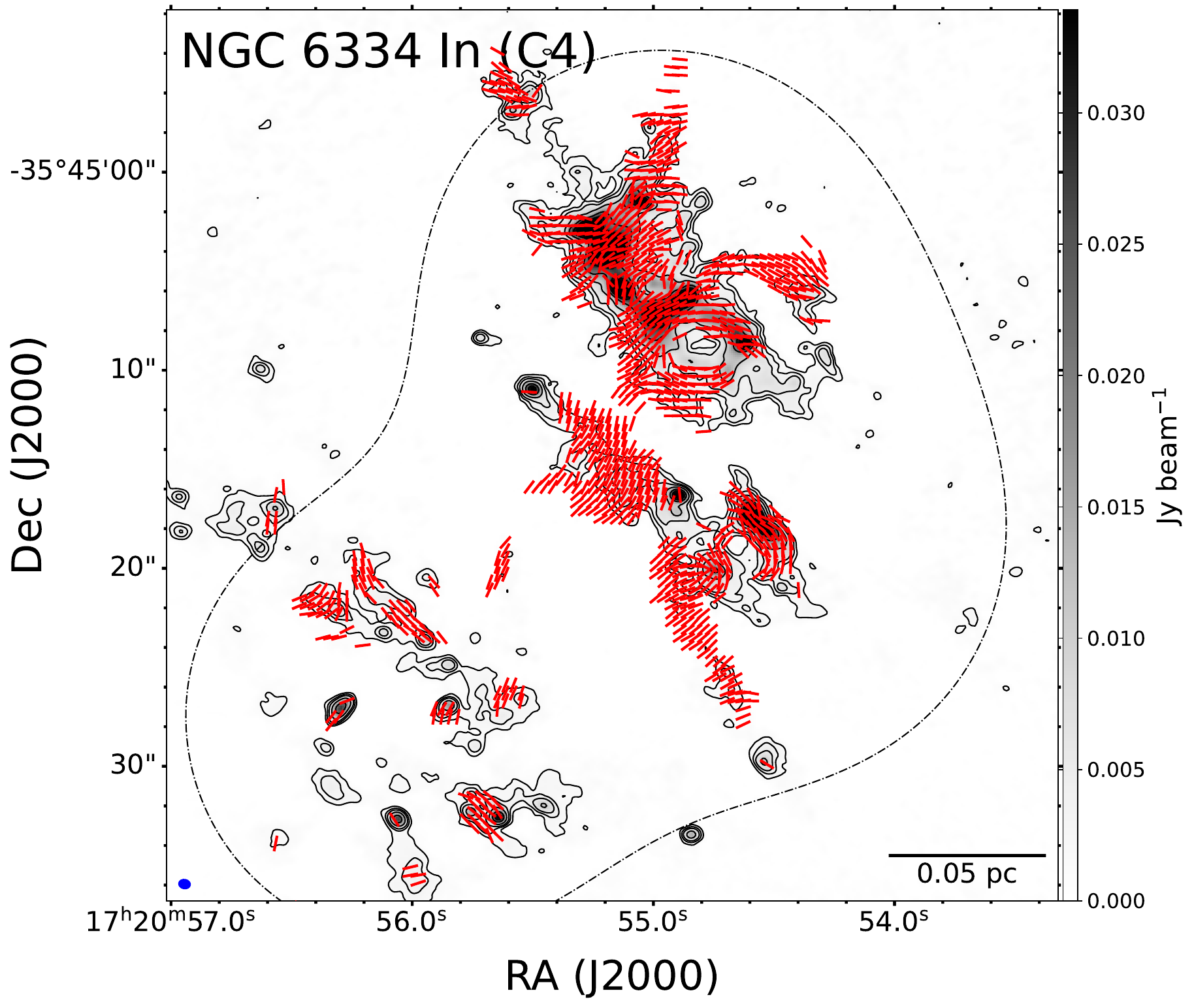}
\includegraphics[width=0.32\textwidth]{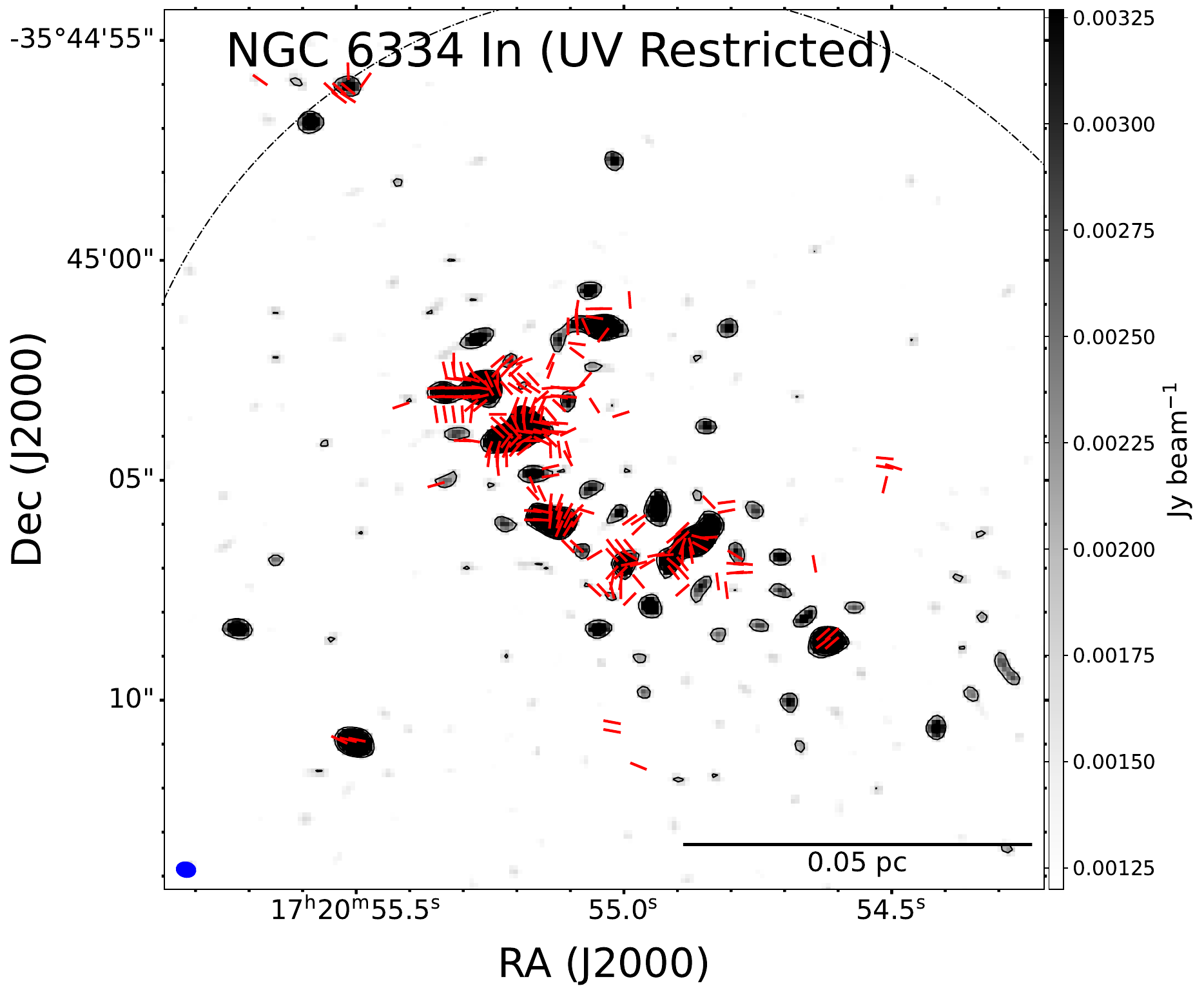}
\caption{Magnetic field orientations in NGC 6334 In made from the C43-1, C43-4, and C43-4-restricted data that exclude visibilities with projected baseline lengths $<$ 250m. The Stokes $I$ emission is shown in contours and grey scales. The red line segments represent the position angles (PA) of the plane-of-the-sky component of magnetic fields. The synthesized beam is marked at the lower-left corner of each panel. The horizontal bar at the lower-right corner of each panel represent a linear scale of 0.05 pc. The dotted line marks the 50\% sensitivity level of the observations.}
\label{fig:N6336In}
\end{figure}

\begin{figure}[!ht]
\centering 
\includegraphics[width=0.9\textwidth]{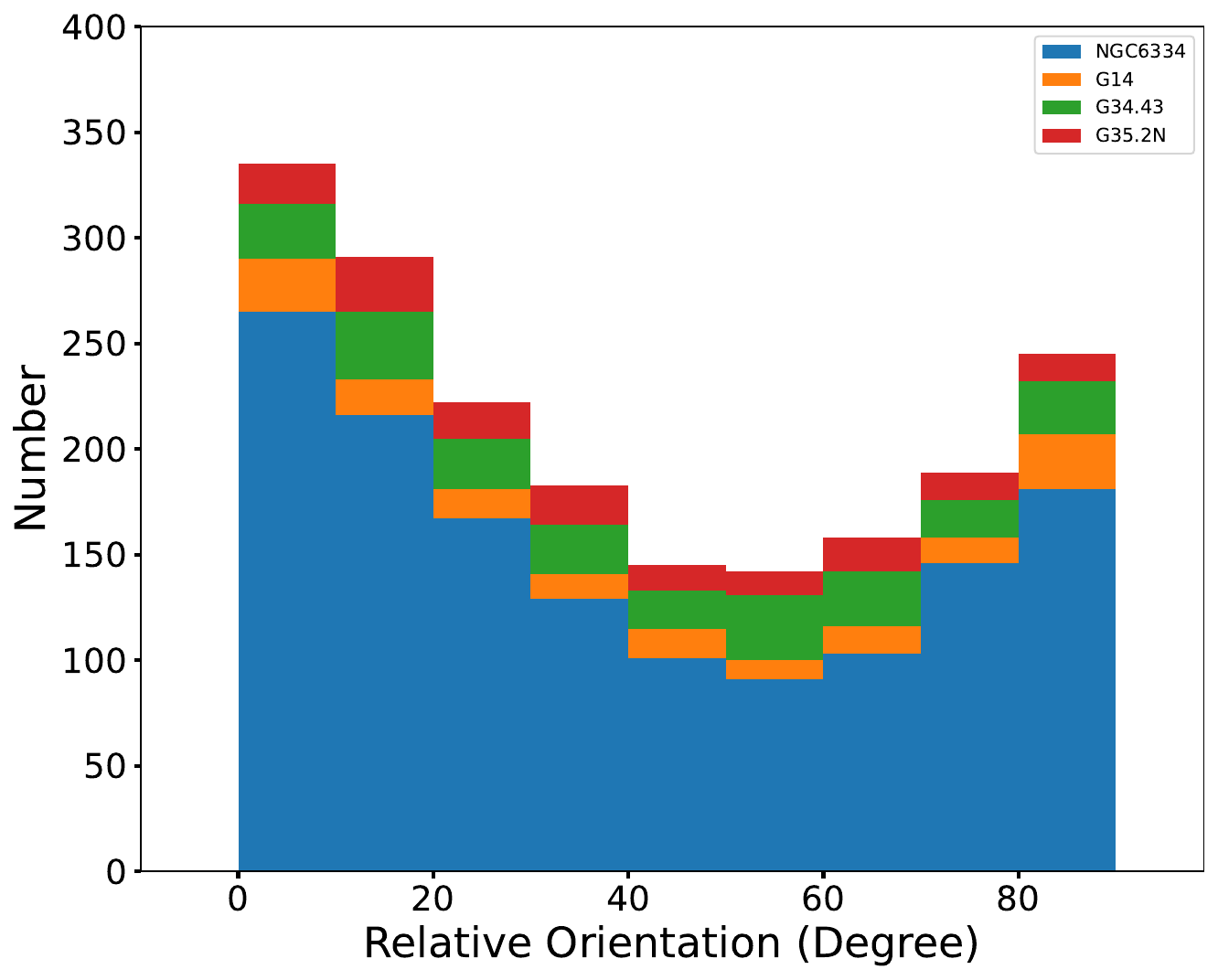}
\caption{Relative orientations of magnetic field position angles between the C43-1 and the C43-4-restricted data that excludes visibilities with projected baselines $<$ 250m. The histogram includes data from NGC 6334, G14, G34.43 and G35.2N.}
\label{fig:angDiff_All}
\end{figure}

\begin{figure}[!ht]
\centering 
\includegraphics[width=0.49\textwidth]{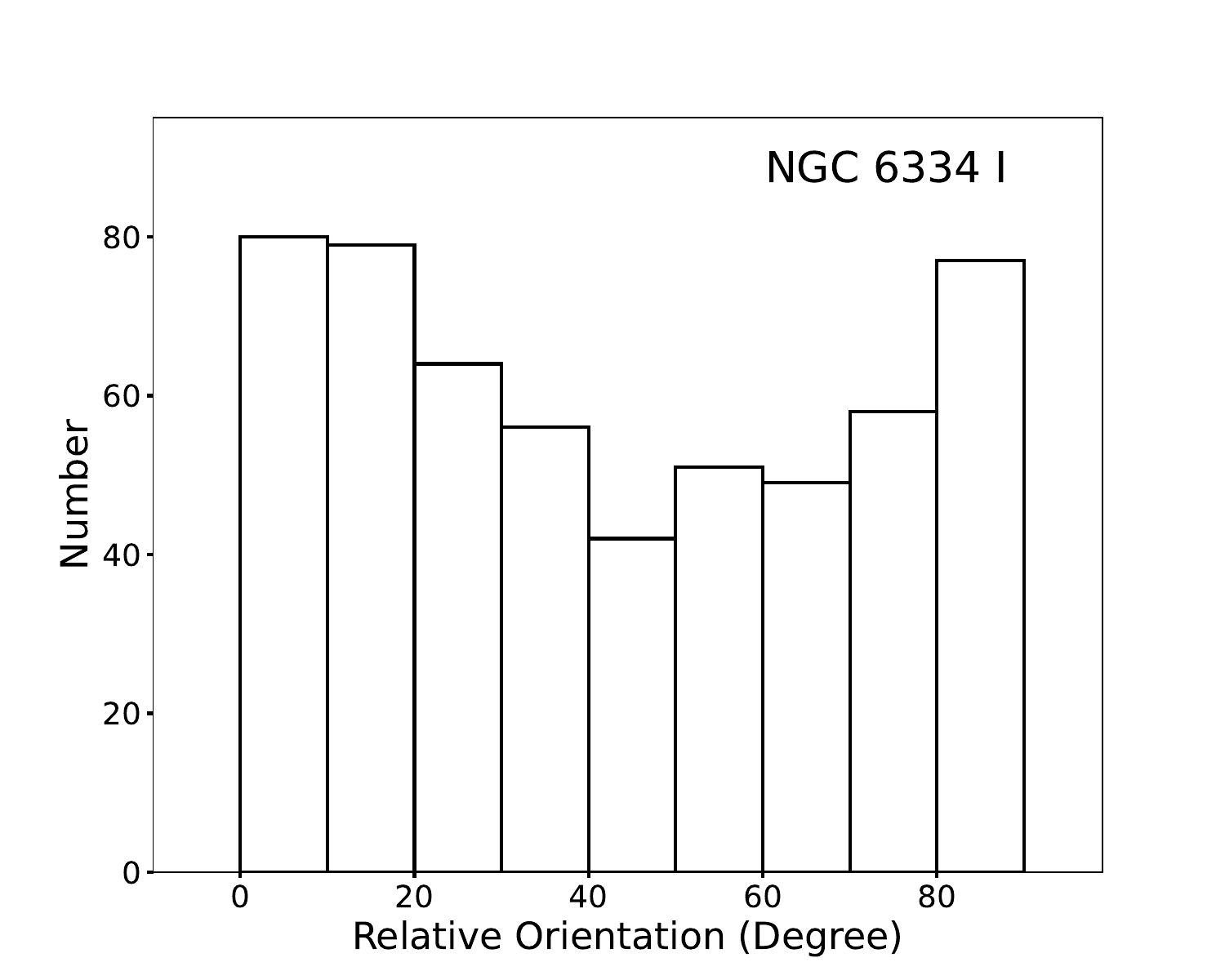}
\includegraphics[width=0.49\textwidth]{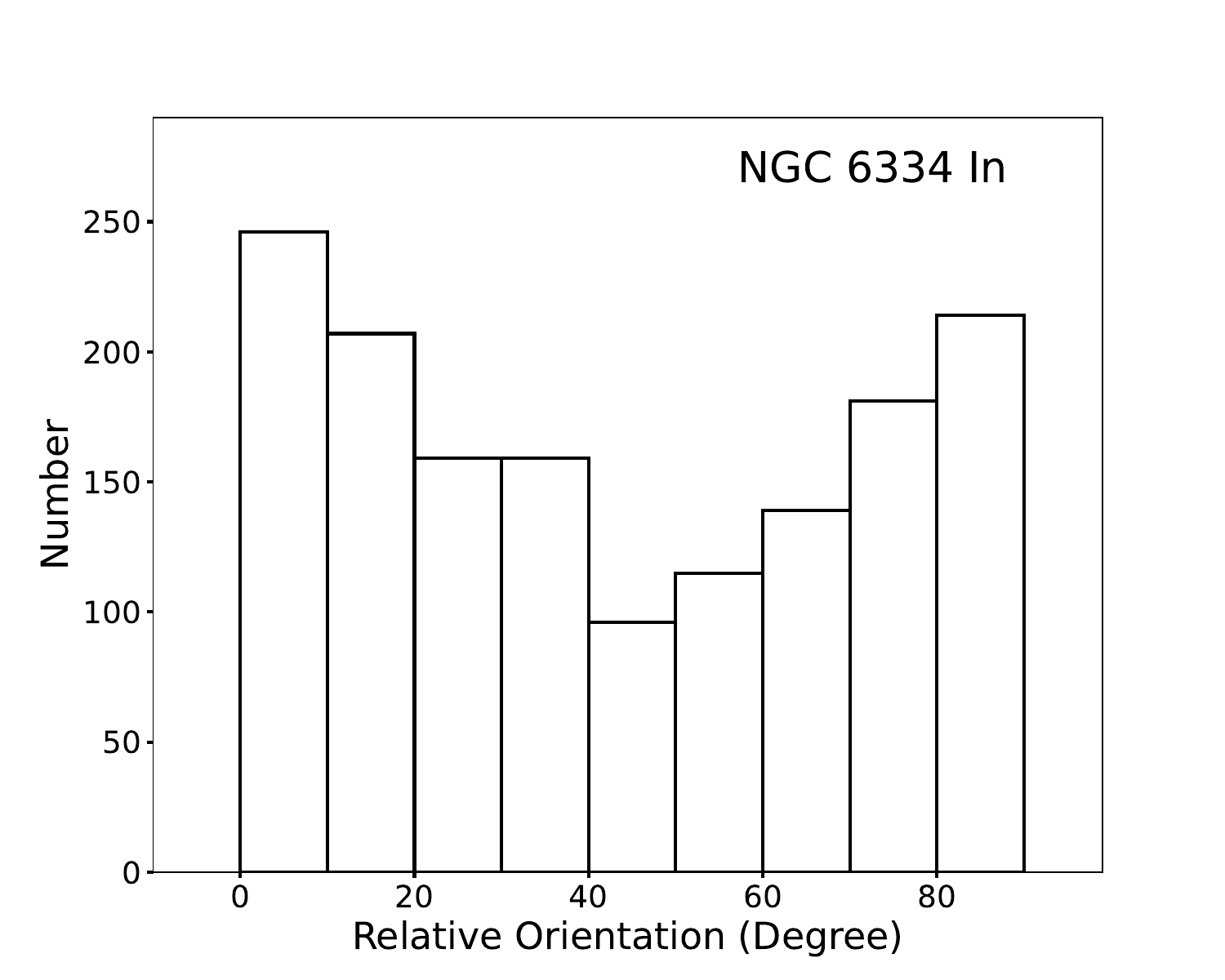}
\includegraphics[width=0.49\textwidth]{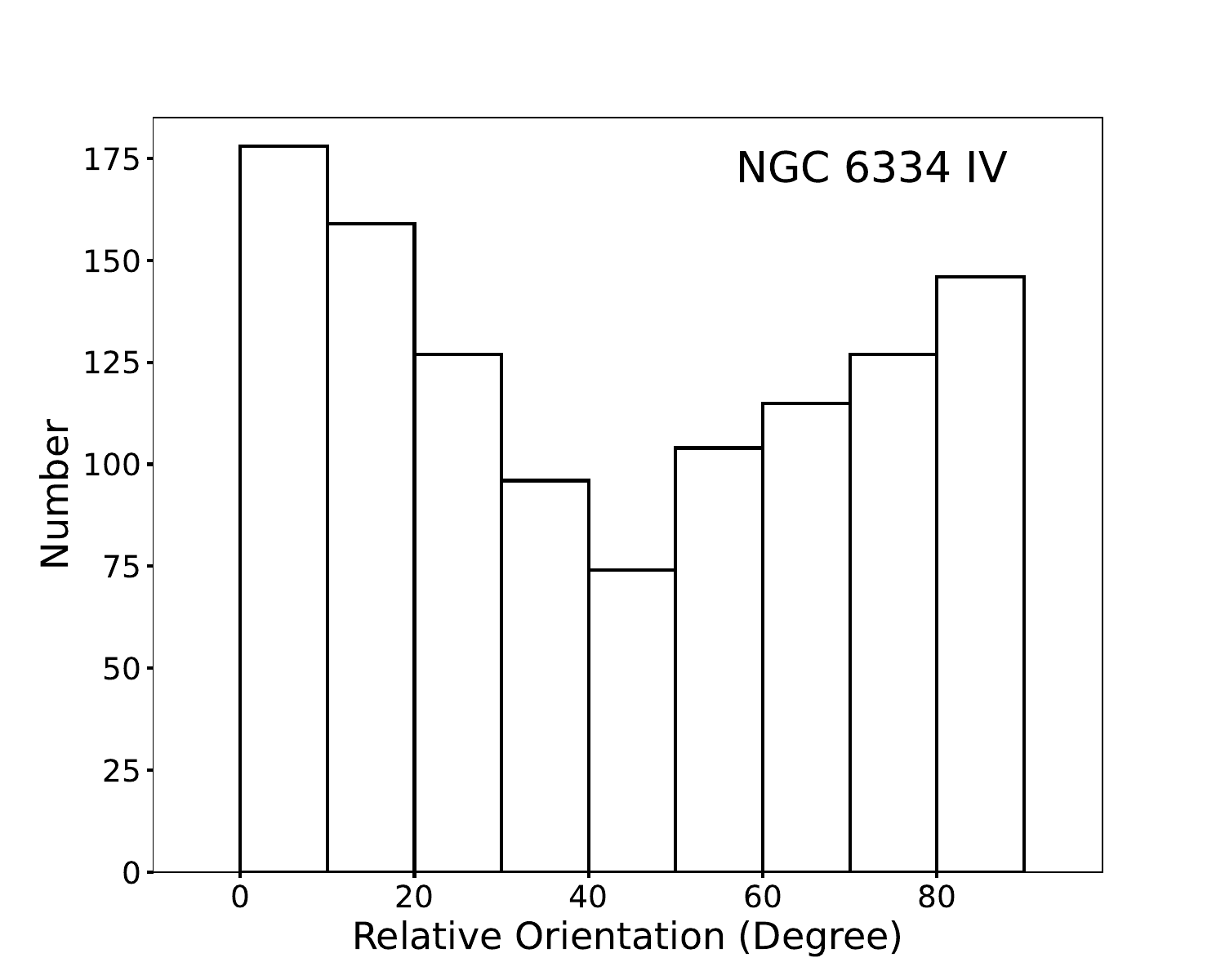}
\includegraphics[width=0.49\textwidth]{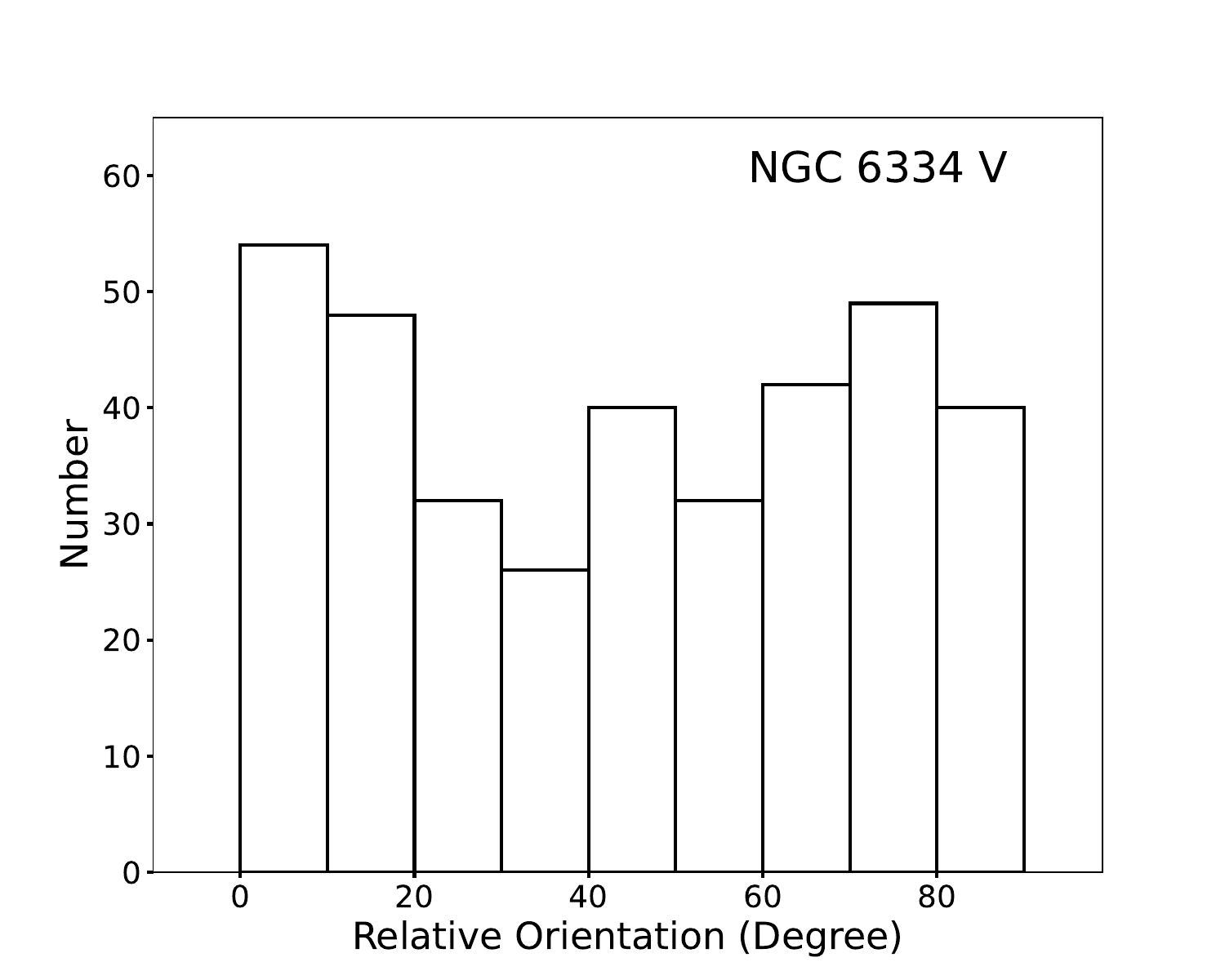}
\caption{Relative orientations of magnetic field position angles between the C43-1 and the C43-4-restricted data that excludes visibilities with UV distances $<$ 250m. }
\label{fig:angDiff_N6336I}
\end{figure}

\begin{figure}[!ht]
\centering 
\includegraphics[width=0.49\textwidth]{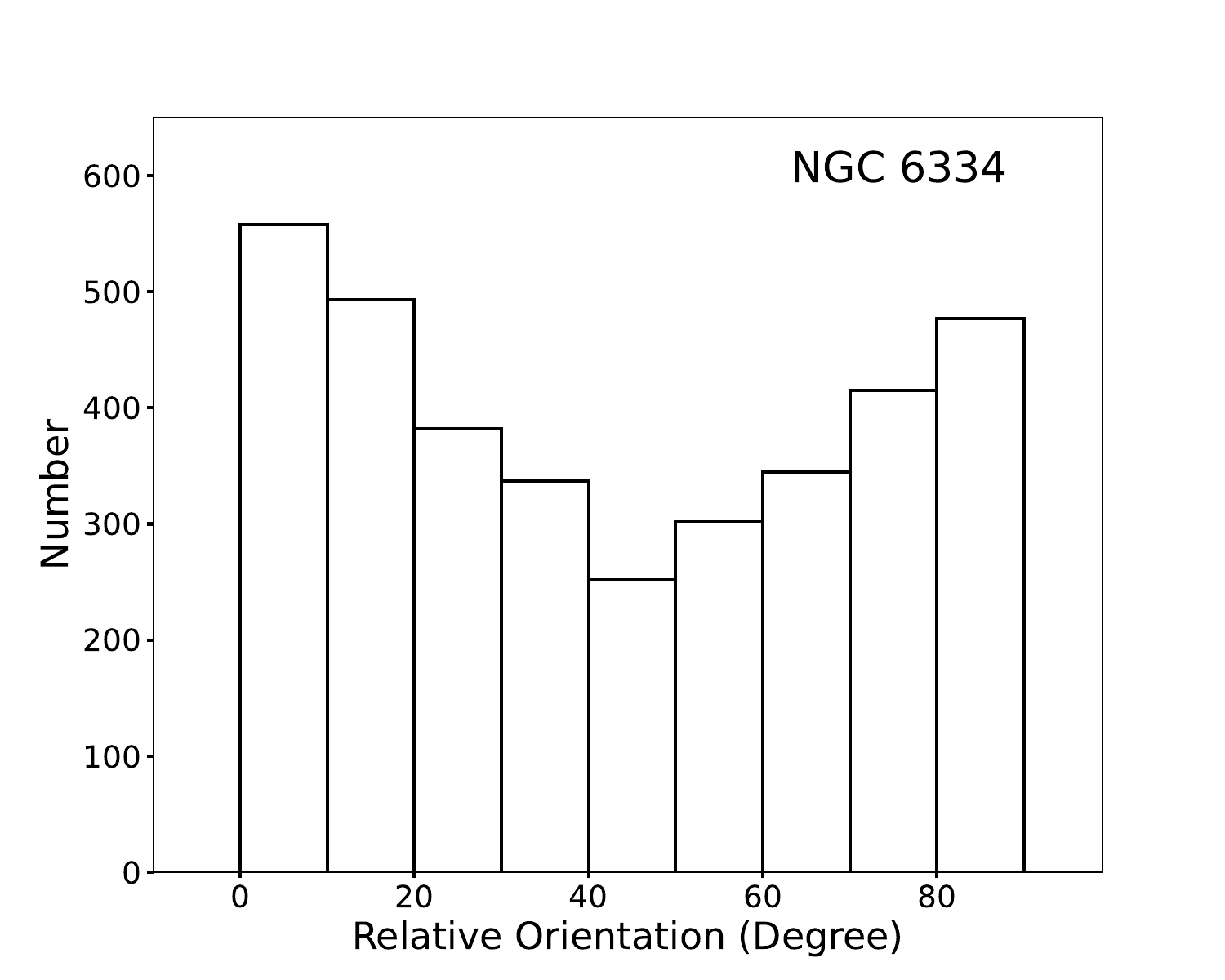}
\includegraphics[width=0.49\textwidth]{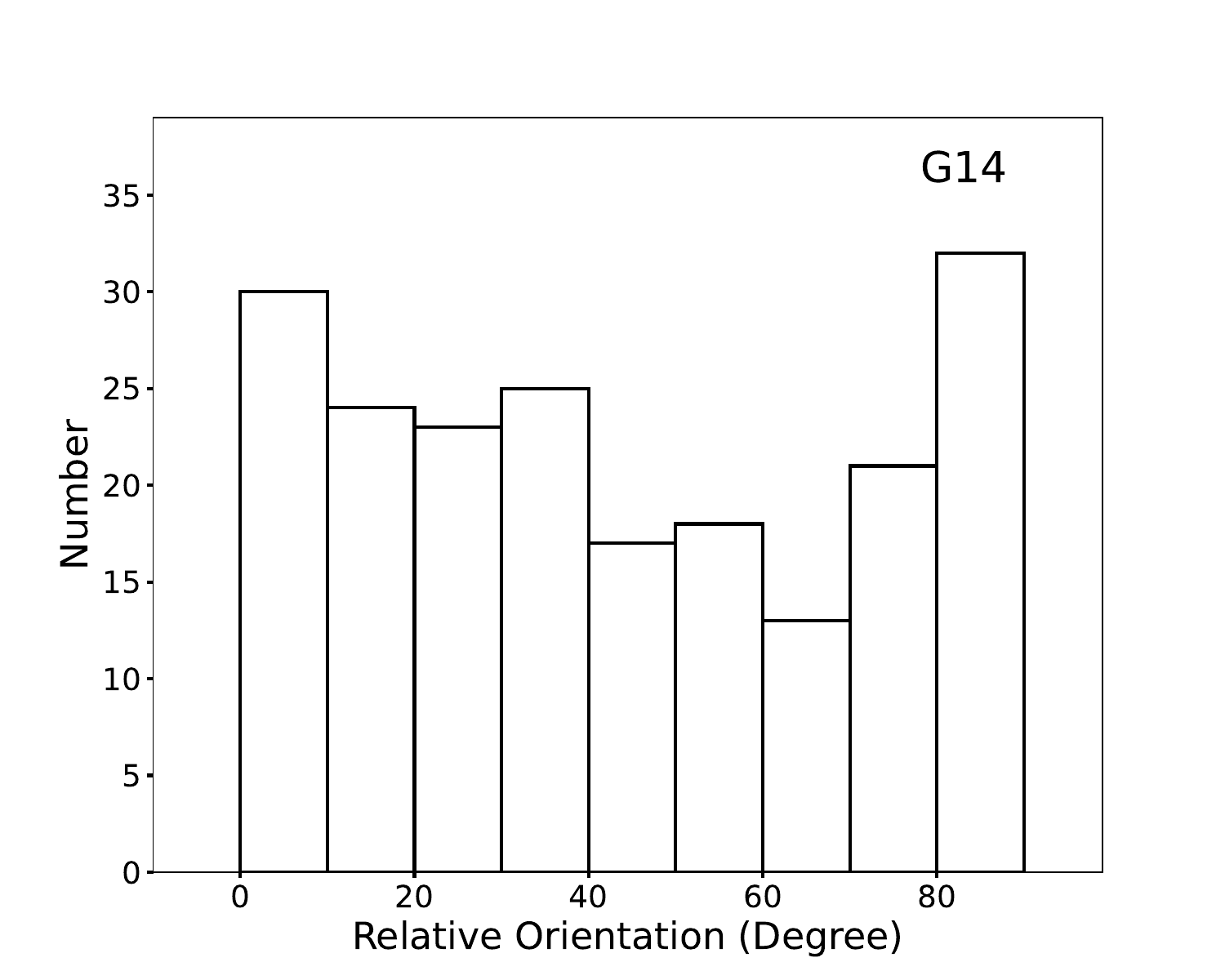}
\caption{Relative orientations of magnetic field position angles between the C43-1 and the C43-4 data excluding visibilities with UV distances $<$ 250m. The left panel shows the combined RO distribution of NGC~6334 I, In, IV and V. The right panel shows the combined RO distribution of G14.}
\label{fig:angDiff_N6336All}
\end{figure}

\begin{figure}[!ht]
\centering 
\includegraphics[scale=0.9]{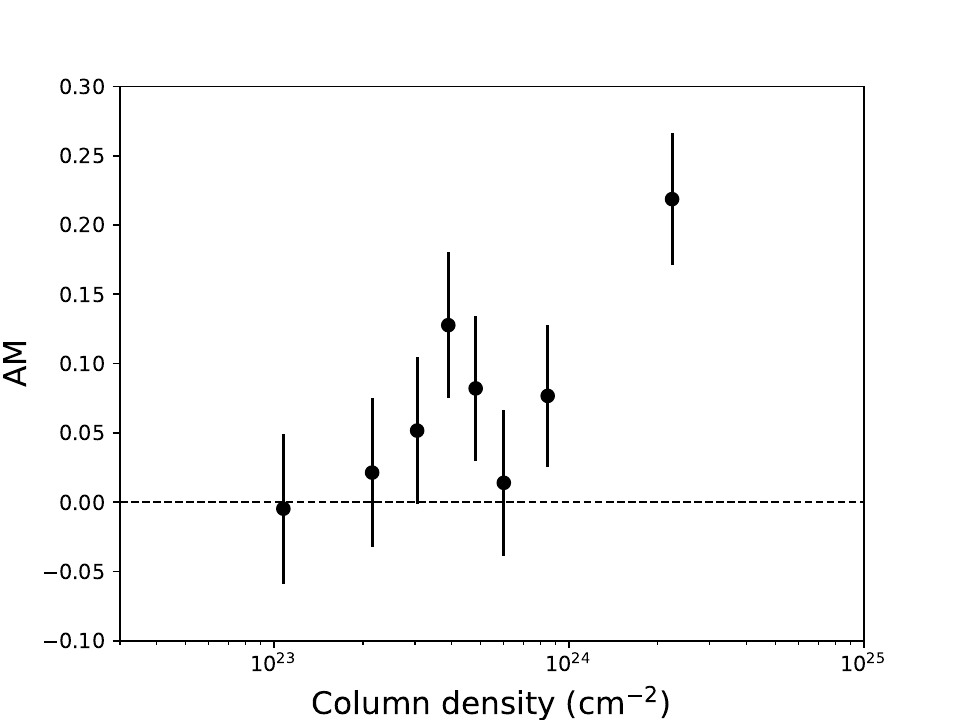}
\caption{Normalized projected Rayleigh statistic of relative orientations of magnetic fields between the C43-1 and the C43-4-restricted data that excludes visibilities with UV distances $<$ 250m versus column densities for NGC~6334. }
\label{fig:N6334all_PRS}
\end{figure}

\begin{figure}[!ht]
\centering 
\includegraphics[width=0.9\textwidth]{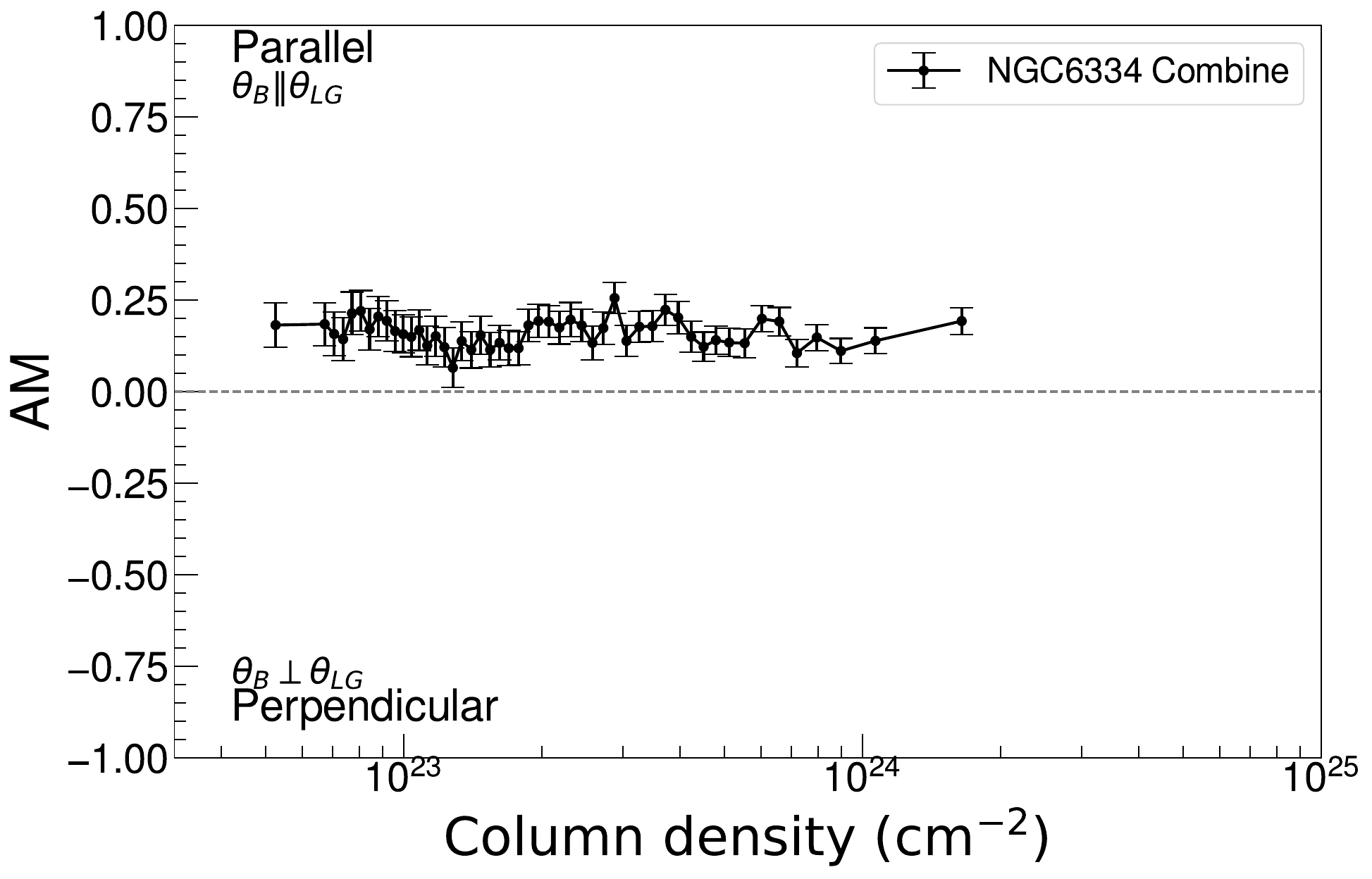}
\caption{Normalized PRS analysis of ROs between directions of local gravity and magnetic fields for NGC~6334. }
\label{fig:N6334all_PRS_LG}
\end{figure}

\begin{table*}
\caption{List of Observations }
\label{tab:1}
\begin{center}
\begin{tabular}{l|ccccc}
\hline \hline
       Date of      &  Array        & Flux/Gain/BP/Polarization$^{a}$ & Targets \\
    Observations & Configuration & Calibrators   & --   \\
\hline
2018-03-31    &     C43-4   &  J1924/J1832/J1924/J1924  & IRDC G14      \\ 
2018-07-17    &     C43-1   &  J1924/J1832/J1924/J1924  & IRDC G14      \\ 
2018-09-06    &     C43-4   &  J1924/J1832/J1924/J1924  & IRDC G14         \\ 
2018-11-09    &     C43-4   &  J1924/J1832/J1924/J1924  & IRDC G14     \\ \hline
2018-06-23    &     C43-1   &  J1751/J1851/J1751/J1924  &  G28/G34.41/I18360/G35      \\ 
2018-06-29    &     C43-1   &  J1751/J1851/J1751/J1924  &  G28/G34.41/I18360/G35      \\ 
2018-09-09    &     C43-4   &  J1751/J1851/J1751/J1924  &  G28/G34.41/I18360/G35      \\ 
2018-09-10    &     C43-4   &  J1751/J1851/J1751/J1924  &  G28/G34.41/I18360/G35      \\ 
2018-09-12    &     C43-4   &  J1751/J1851/J1751/J1924  &  G28/G34.41/I18360/G35     \\ 
2018-09-13    &     C43-4   &  J1751/J1851/J1751/J1924  &  G28/G34.41/I18360/G35      \\ \hline
2018-06-28    &     C43-1   &  J1924/J1733/J1924/J1924  &  NGC6334       \\ 
2018-09-03    &     C43-4   &  J1924/J1733/J1924/J1924  &  NGC6334       \\ \hline
\end{tabular}
\tablenotetext{a}{Calibrators J1733, J1751, J1851/ and J1924 stand for J1733-3722, J1751+0939, J1851+0035, J1832-2039, J1924-2914, respectively}
\end{center}
\end{table*}

\begin{table*}
\caption{Sample of Massive Star Forming Regions}
\label{tab:2}
\begin{center}
\begin{tabular}{l|rr|c|c|C|l}
\hline
Source & $\alpha$ (J2000) & $\delta$ (J2000) & $V_{\rm lsr}$  & D  & Beam & Image rms (mJy)$^a$ \\
(pointing) &    hh:mm:ss  &  dd:mm:ss     &  (km s$^{-1}$) &   (kpc) &  $^{\prime\prime} \times ^{\prime\prime}$ &  I/Q/U \\
\hline
NGC6334I      & 17:20:53.41  & -35:46:57.8 & -8 &  1.3 & 0.68 $\times$ 0.52 & 2.0/0.057/0.053 \\
\hline
NGC6334In (1) & 17:20:54.97  & -35:45:05.6 & -8 &  1.3 &  &\\
NGC6334In (2) & 17:20:54.53  & -35:45:18.8 & -8 &  1.3 & 0.80 $\times$ 0.60 & 0.60/0.048/0.052 \\
NGC6334In (3) & 17:20:56.00  & -35:45:27.5 & -8 &  1.3 &  & \\
\hline
NGC6334IV (1) & 17:20:54.97  & -35:45:05.6 & -8 &  1.3 &  & \\
NGC6334IV (2) & 17:20:18.24  & -35:54:42.7 & -8 & 1.3 & 0.79 $\times$ 0.59 & 0.66/0.048/0.045 \\
NGC6334IV (3) & 17:20:18.19  & -35:54:52.7 & -8 & 1.3 &   & \\
\hline
NGC6334V      & 17:19:57.55  & -35:57:50.8 & -8 & 1.3 & 0.69 $\times$ 0.53 & 0.47/0.046/0.048 \\
\hline
NGC6334IR (1)      & 17:19:08.79  & -36.07.03.1 & -8 & 1.3 &  & \\
NGC6334IR (2)      & 17:19:07.47  & -36.07.15.0 & -8 & 1.3 & 0.80 $\times$ 0.59 & 0.079/0.024/0.024 \\
NGC6334IR (3)      & 17:19:05.92  & -36.07.21.4 & -8 & 1.3 &  & \\
\hline
G14S (1)      & 18:18:13.80  & -16:57:11.5 & 20 & 2.3 &  & \\
G14S (2)      & 18:18:13.08  & -16:57:21.6 & 20 & 2.3 &  & \\
G14S (3)      & 18:18:12.39  & -16:57:22.4 & 20 & 2.3 & 0.70 $\times$ 0.61 & 0.030/0.017/0.017 \\
G14S (4)      & 18:18:11.36  & -16:57:28.5 & 20 & 2.3 &  & \\
\hline
G14N (1)      & 18:18:12.44  & -16:49:27.1 & 20 & 2.3 &  &  \\
G14N (2)      & 18:18:13.02  & -16:49:40.2 & 20 & 2.3 & 0.71 $\times$ 0.62 & 0.15/0.021/0.021 \\
G14N (3)      & 18:18:12.57  & -16:50:08.6 & 20 & 2.3 &  &  \\ \hline
G14N (4)      & 18:18:11.34  & -16:52:34.0 & 20 & 2.3 & 0.55 $\times$ 0.54 & 0.060/0.021/0.022   \\ \hline
G14N (5)      & 18:18:06.84  & -16:51:19.4 & 20 & 2.3 & 0.55 $\times$ 0.54 & 0.036/0.021/0.021 \\ \hline
G14N (6)      & 18:18:11.50  & -16:51:35.7 & 20 & 2.3 & 0.55 $\times$ 0.54 & 0.030/0.021/0.021 \\ \hline
I18360        & 18:38:40.73  & -05:35:04.0 & 103.5 & 6.3 & 0.44 $\times$ 0.35 & 0.43/0.040/0.040 \\
\hline
G28-MM4 (1)  & 18:42:51.05  &  -04:03:08.6  &  78.0  &  4.8  & & \\
G28-MM4 (2)  & 18:42:50.69  &  -04:03:13.4  &  78.0  &  4.8  & & \\
G28-MM4 (3)  & 18:42:50.39  &  -04:03:19.0  &  78.0  &  4.8  & 0.67 $\times$ 0.51 & 0.020/0.011/0.010  \\
G28-MM4 (4)  & 18:42:49.88  &  -04:03:24.47  &  78.0  &  4.8  &  & \\ \hline
G28-MM9 (1)  & 18:42:46.48  &  -04:04:14.27  &  78.0  &  4.8  & 0.67 $\times$ 0.50 & 0.020/0.013/0.013 \\ 
G28-MM9 (2)  & 18:42:46.91  &  -04:04:09.34  &  78.0  &  4.8  &  & \\
\hline
G34.43 (1)      & 18:53:18.01 &  01:25:25.6 & 57 & 1.57 & 0.45 $\times$ 0.36 & 0.35/0.037/0.037 \\ \hline
G34.43 (2)      & 18:53:18.44 &  01:24:54.1 & 57 & 1.57 & 0.50 $\times$ 0.40 & 0.10/0.024/0.024 \\
G34.43 (3)      & 18:53:18.67 &  01:24:41.2 & 57 & 1.57 &  & \\
\hline
G35.2N (1)     & 18:58:12.72 &  01:40:42.7 & 34 & 2.19 &  & \\
G35.2N (2)     & 18:58:13.08 &  01:40:33.6 & 34 & 2.19 & 0.50 $\times$ 0.40 & 0.30/0.030/0.03 \\ \hline
G35.2N (3)     & 18:53:20.40 &  01:28:23.0 & 34 & 2.19 & 0.45 $\times$ 0.36 & 0.048/0.026/0.026 \\
\hline
\end{tabular}
\tablenotetext{a}{Values represent the $1\sigma$ rms noise level in images from the combined data from array configurations C43-1 and C43-4.}
\end{center}
\end{table*}

\bibliography{bibliography.bib}

\end{document}